\documentclass[seceq]{ptptex}
\usepackage{amsmath,amssymb,color,amssymb,latexsym,graphicx,
mathrsfs,
}

\newcommand{\ma}[1]{\mbox{$\mathcal{#1}$}}

\newcommand{\D}{{\rm d}}
\newcommand{\we}{\wedge}

\title{
Black hole solutions in string theory%
}

\author{
Kei-ichi \textsc{Maeda}$^{1,2}$ and Masato \textsc{Nozawa}$^{3}$%
}

\inst{
$^1$Department of Physics, 
Waseda University, Tokyo 169-8555, Japan, \\
$^2$Research Institute for Science and Engineering,Waseda University,
 Tokyo 169-8555, Japan, \\
$^3$Theory Center, IPNS, KEK, 1-1, Oho, Tsukuba, Ibaraki, 
305-0801, Japan}

\abst{
Supersymmetric solutions of supergravity have been of particular
importance in the advances of string theory.  
This article reviews the current status of black hole solutions 
in higher-dimensional supergravity theories. 
We discuss primarily the gravitational aspects of supersymmetric 
black holes and their relatives in various dimensions. 
Supersymmetric solutions and their systematic derivation 
are reviewed with prime examples.
We also study the stationary or dynamically intersecting branes 
in ten and eleven-dimensions, which provide a number of interesting black objects 
via the dimensional reduction and duality transformations.  
}

\begin{document}

\maketitle

\vspace{-8.0cm}
\hfill{KEK-TH 1453}
\vspace{7.5cm}

\tableofcontents

\section{Introduction}

Gravitational physics in higher dimensions has 
played and will continue to play a central r\^ole 
in the development of string theory as well as
phenomenological predictions of brane world scenarios.  
Among other things, a various kinds of interesting 
higher-dimensional ``black objects''
have been found and extensively studied over the last two decades.  
These black objects exhibit much richer physical properties--even in the
vacuum case--than the four-dimensional counterparts\cite{supplement}. 
One of the most exciting findings is the discovery of a black ring~\cite{ER,PS}, 
which describes an asymptotically flat regular black ``hole'' with 
topology $S^1\times S^2$.   Since there also exists an ordinary rotating 
spherical black hole~\cite{MP} with the same mass and angular
momentum as the ring in a certain range of parameters,  it exemplified the 
end of black-hole uniqueness in higher dimensions and have triggered 
intensive works on higher dimensional gravity\cite{supplement}. 
Another novel phenomenon intrinsic to higher dimensions is the 
dynamical gravitational instability of black branes~\cite{GL,GL2}. 
This instability provides a considerably rich phase diagram 
with new insights, such as non-uniqueness,  topology changing 
phase transitions between black objects~\cite{BHBS_transition,
BHBS_transition2} and new kind of black holes with pinched 
horizons~\cite{Emparan:2007wm}.

Brane configurations in supergravity are much more relevant 
to string theory as a low energy description of D-branes. 
Since these black branes preserve some fraction of supersymmetries, 
we are able to bring  their behavior in the strongly
coupled regime under some control. Supersymmetric configurations    
often circumvent instability and have non-renormalization
properties~\cite{Witten:1978mh}, which are not applicable for
non-supersymmetric ones.
A principal achievement of the study of
supersymmetric black holes is the microscopic derivation of 
Bekenstein-Hawking entropy~\cite{Bekenstein:1973ur,Hawking:1974sw} exploiting the D-brane 
technology~\cite{Strominger:1996sh}. 
String theory promoted the conjecture of Bekenstein about 
the striking resemblance between black-hole mechanics and
ordinary thermodynamics~\cite{Bekenstein:1973ur,Bardeen:1973gs} to more than an analogy. 
A number of subsequent developments since the pioneering work of
Strominger \& Vafa~\cite{Strominger:1996sh} 
have witnessed a significant progress in the statistical description of
black hole entropy  and highlighted 
the utility of supersymmetric intersecting brane configurations (see
Ref.~\cite{Maldacena:1996ky} for a review).

In this article, we shall review black holes in supergravity theories. 
Since these studies are extremely broad in scope 
ranging from the phenomenological topic to the basis of
AdS/CFT-correspondence~\cite{Maldacena:1997re}, 
it is almost impossible to cover all aspects. 
So we shall stick to the discussion of gravitational
physics of various higher-dimensional supersymmetric black holes 
and their cousins, with an emphasis on the distinction 
from four-dimensional ones and from the non-supersymmetric ones.

The organization of the rest of the paper is as follows. 
In the next section we shall make a brief overview of supersymmetric black
holes in four dimensions. Section~\ref{sec:classifications} is devoted to
the  classification of supersymmetric solutions. 
Focusing mainly on the five dimensions, we will study some black hole
solutions. In section~\ref{sec:intersecting_branes}, 
the intersecting black brane solutions with rotations are discussed,
which produce upon dimensional reduction various kinds of lower dimensional black holes. 
The extension to the dynamical background is examined in
section~\ref{sec:dynamical_branes}. It is shown that 
the dimensional reduction of these 
time-dependent branes provides black holes in an expanding
Friedmann-Lema\^itle-Robertson-Walker (FLRW) universe, on which 
the standard cosmological evolutionary scenario is based. 
Section~\ref{sec:conclusions} concludes with several remarks.

We shall work with mostly plus metric signature.
The convention of the Dirac matrix $\gamma_\mu $ 
is $\bar \psi :={\rm i}\gamma^0 \psi^\dagger $. 
The four-dimensional chiral matrix is  given by 
$\gamma_5=i \gamma_{0123}$.

\section{Supersymmetric black holes: a retrospective}
\label{sec:prologue}

\subsection{A Bogomol'nyi bound}
 
Since the gravitational energy is negative 
due to the universally attractive nature of gravity,   
it is not {\it a priori} clear whether the total energy of the system is positive. 
Thus, the establishment of the positive energy theorem
is one of the most sensational breakthroughs in mathematical
theory of relativity. Stated precisely, the 
Arnowitt-Deser-Misner (ADM) energy~\cite{ADM}
of an asymptotically flat spacetime must be non-negative 
under the dominant energy condition, and vanishes  
only for the flat spacetime. 
The first proof given by Schoen and 
Yau~\cite{Schon:1979rg} is based on the variations of the extremal 
surfaces.\footnote{
The proof of Schoen-Yau~\cite{Schon:1979rg} limits its validity to $D\le 8$
dimensions: otherwise the regularity of minimal surfaces breaks down. 
This might be related to the failure of Bernstein's conjecture in $D\ge 8$
dimensions~\cite{Gibbons:2009ye}. 
By contrast, Witten's argument works in any dimensions 
as far as the spin manifold is concerned.} 
A completely different but extremely simple proof was
given by Witten~\cite{Witten:1981mf} and later refined by Nester~\cite{Nester}. 
A notable feature of Witten's argument is that the ADM momentum is
expressed in terms of an auxially spinor field $\epsilon $. The use of the
spinor field was originally motivated by an attempt to construct
quantum theory of supergravity.  Nevertheless, 
Gibbons and Hull~\cite{Gibbons:1982fy} demonstrated that 
its origin can be found within the classical regime of supergravity.  
They considered the four-dimensional $N=2$ supergravity, whose bosonic
sector is  the  Einstein-Maxwell theory, and obtained a stronger inequality
\begin{align}
 M\ge \sqrt{Q^2+P^2 } \,,
\label{BPS_bound}
\end{align}   
where $M$ is the ADM energy, 
$Q$ and $P$ correspond to the total electric and magnetic charges,
respectively. 
In the context of supergravity, 
they enter the supersymmetry algebra as central charges:
\begin{align}
 \{\mathbf Q, \mathbf Q^*\}= i\gamma^\mu C P_\mu+(Q+i\gamma_5 P)C \,, 
\label{super_algebra}
\end{align} 
where $C$ is the charge conjugation matrix, $\mathbf Q$ is the Dirac
supercharge and $P_\mu $ is the ADM four momentum with 
$M=\sqrt{-P^\mu P_\mu }$. 
The ADM energy and the electromagnetic charges are all defined by
surface integrals at infinity. The above equation~(\ref{super_algebra}) manifests that 
the Witten-Nester expression of ADM energy is obtained by 
the repeated supersymmetry transformations
generated by an associated Noether charge.~\cite{Horowitz:1982mu}.

The inequality~(\ref{BPS_bound}) is reminiscent of the 
Bogomol'nyi-Prasad-Sommerfield (BPS) bound
for solitonic objects.  In the context of gauge theory, 
it is known that the solitonic configuration saturating this bound
fulfills a certain kind of 1st-order differential equations.  
Furthermore multiple configurations of solitons are allowed. 
This is attributed to a delicate balance of forces between the 
electromagnetic repulsion and the attractive force caused by other fields, 
e.g., a scalar field.  
The BPS objects are stable classically, and  
in the supersymmetric case
they are even stable quantum mechanically~\cite{Witten:1978mh}. 
The number of BPS states does not change if 
coupling constants are varied continuously, since it is essentially a 
topological invariant quantity. This is the key ingredient characterizing
the non-perturbative nature of BPS states.

The inequality~(\ref{BPS_bound}) implies that the 
same argument carries over to the gravitating system. 
It is known that the lower bound is attained, or equivalently the matrix
$\{\mathbf Q, \mathbf Q^* \}$ is not of the maximal rank, 
 if and only if the asymptotically flat spacetime 
admits a spinor $\epsilon $ satisfying the analogue of 
first-order BPS equation~\cite{Gibbons:1982fy}
\begin{align}
\hat \nabla_\mu \epsilon:= \left(\nabla_\mu 
+\frac{i}{4}F_{\nu\rho }\gamma^{\nu\rho }\gamma_\mu \right)\epsilon =0
 \,.
\label{susy} 
\end{align}
This can be viewed as the vanishing of gravitino supersymmetry
transformation 
$\delta_\epsilon  \psi _\mu =\hat \nabla_\mu \epsilon =0$, 
which leaves the bosonic background invariant (note that $\epsilon $
is anti-commuting in this context, whereas it is taken to be commuting 
in the proof of positive energy theorem).  
Since the BPS equation~(\ref{susy}) is the 1st-order linear
differential equation, supersymmetric solutions 
are very simple and expected to inherit properties of instantons.

It is worthwhile to note that 
the integrability of the Killing spinor equation allows us to 
relate the BPS geometries admitting a Killing spinor to those
solving equations of motion.  
Acting a supercovariant derivative to Eq.~(\ref{susy}) and anti-symmetrizing
the indices, one obtains 
$\hat \nabla_{[\mu }\hat \nabla_{\nu]}\epsilon =0$.
Contracting this with $\gamma^\nu $,  
using 
$\nabla_{[\mu}\nabla_{\nu]}\epsilon
=(1/8)R_{\mu\nu\rho\sigma}\gamma^{\rho\sigma}\epsilon $ 
and the Bianchi identity $R_{\mu[\nu\rho\sigma] }=0$, 
it is a simple exercise to obtain
\begin{eqnarray}
\left[E_{\mu\nu }\gamma^\nu +\frac{i}{2}\left(\gamma^{\nu\rho\sigma }
\gamma_\mu 
\nabla_{[\nu}F_{\rho\sigma]}-2\gamma_\nu\gamma_\mu \nabla_\rho
F^{\nu\rho }\right)\right] \epsilon=0\,, 
\label{KS_integrability}
\end{eqnarray}
where 
$E_{\mu\nu }:=R_{\mu\nu }-2(F_{\mu\rho }{F_\nu}^\rho -\tfrac 14F^2g_{\mu\nu })$ 
and its vanishing is equivalent to the Einstein equations. 
If the Maxwell equations and the Bianchi identity for the Maxwell field
are satisfied, the 
last two terms in Eq.~(\ref{KS_integrability}) drop off, giving 
$E_{\mu\nu }\gamma^\nu \epsilon =0$. 
Contracting this with $\bar \epsilon $ and $E_{\mu\rho }\gamma^\rho $, 
we obtain
\begin{align}
 E_{\mu\nu }V^\nu  =0 \,, \qquad 
 E_{\mu\nu }{E_\mu }^\nu =0 \quad (\textrm{no sum on $ \mu $}). 
\end{align}
It is clear that $V^\mu =i \bar \epsilon \gamma^\mu \epsilon $
cannot be spacelike since we have $V^0 =\epsilon^\dagger \epsilon >0$
for a nonvanishing spinor $\epsilon $ satisfying Eq.~(\ref{susy}). 
In the case where $V^\mu =i\bar\epsilon\gamma^\mu \epsilon $
is timelike,  the first equation means that 
$E_{00}=E_{0i}=0$ and the second equation implies that 
$E_{ij}=0$ (in an obvious orthonormal frame), as we desired. 
In the null case solving  
only the component $E_{++}=0$ (in the basis $V=e^+$)
suffices to ensure that other components of Einstein's equations 
automatically follow. 
Conversely, 
the integrability condition can be used to check consistency
for the 1st-order equation~\cite{nozawa}.

Although supersymmetric solutions are characterized by 
the BPS relation, there is no general proof that any theories
admit the Bogomol'nyi {\it bound} applied to non-supersymmetric solutions. 
Ref.~\cite{Gibbons:1994vm} investigated circumstances under which   
one can derive the bound ({\it \`a la} Witten-Nester), and 
found that this is the case 
only for the solutions of bosonic sector of supergravity.   
This fact also validates the stability of supersymmetric solutions 
with a fair amount of certainty.

\subsection{Supersymmetric black holes in four dimensions}

An illustrative example of the supersymmetric solution 
is the Majumdar-Papapetrou solution~\cite{Majumdar:1947eu}. 
Let us begin by the electrically charged extremal 
Reissner-Nordstr\"om solution with the ADM mass $M=Q$. 
In the isotropic coordinates, the solution is given by  
\begin{align}
\D s^2_4 =- \left(1+\frac{Q}{r}\right) ^{-2}\D t^2 +
\left(1+\frac{Q}{r}\right)^2 \left(\D r^2+r^2 \D \Omega_2^2 \right)
\,, \qquad 
A=\frac{\D t}{1+Q/r } \,, 
\label{RN}
\end{align}
where $\D \Omega_2^2$ is the line element of a unit two-sphere.
Observing that the term $1+Q/r$ describes a monopole  
harmonic on the Euclid 3-space $\mathbb R^3$, 
it is shown that any harmonic on the flat 3-space 
still solves the Einstein equations, yielding the Majumdar-Papapetrou solution
\begin{align}
\D s_4^2 =-H^{-2}\D t^2+H^2 \D \vec x ^2 \,, \qquad A=H^{-1}\D t\,,
\label{MP}
\end{align}
where $H$ is independent of $t$ and satisfies 
the Laplace equation $\vec \nabla^2 H=0$. 
Here and hereafter, 3-dimensional vector notation will be used for 
quantities on $\mathbb R^3$. The Majumdar-Papapetrou
solution~(\ref{MP}) is indeed supersymmetric since it admits a nonzero Killing spinor 
$\epsilon =H^{-1/2}\epsilon_\infty$ satisfying Eq.~(\ref{susy}), where $\epsilon_\infty $ is a 
constant spinor satisfying $i\gamma^0 \epsilon_\infty =\epsilon_\infty$.
Since the matrix $\gamma^0$ has eigenvalues $\pm i$ coming in pairs, 
this projection breaks half of the supersymmetries of vacuum. 
For the multi-centered harmonics,  
\begin{align}
 H=1+\sum_{k=1}^N \frac{Q_k}{|\vec x-\vec x_k|} \,, 
\end{align}
the solution describes $N$ charged black holes in static
equilibrium~\cite{Hartle:1972ya},  where 
$Q_k$ represents the individual charge of the point mass $M_k=Q_k$ 
positioned at $\vec x=\vec x_k$. 
The ADM mass is equal to the total charge $Q=\sum_k Q_k$, 
which saturates the BPS bound~(\ref{BPS_bound}) as expected.

One may follow the same steps for the rotating spacetime.  
Let us consider the Kerr-Newman solution with $M=Q$, 
the metric of which reads
\begin{align}
\D s_4^2 =&- \frac{\rho^2}{\Sigma ^2 }
\left[\D t +\frac{aQ(2 r+Q)\sin^2\theta }{\rho ^2}
\D \phi \right]^2 
+\frac{\Sigma ^2 }{\rho^2 }
\left[\rho^2 \left(\frac{\D r^2}{\Delta }+ \D \theta^2 \right)
+\Delta  \sin^2\theta
\D \phi^2 \right]\,,
\label{KN}
\end{align}
with
\begin{align}
\Sigma^2 =(r+Q)^2+a^2\cos^2\theta  \,, \quad 
\Delta =r^2+a^2 \,, \quad 
\rho^2=r^2+a^2\cos^2\theta\,,
\end{align} 
where the radial coordinate $r$ has been shifted by $Q$
from the conventional Boyer-Lindquist coordinates. 
Since $g^{rr}>0$ for $a\ne 0$,  the rotating spacetime~(\ref{KN}) describes 
a naked singularity rather than a black hole. 
The cosmic censorship bound does
not necessarily coincide with the Bogomol'nyi bound. 
Nonetheless,   
this metric is still supersymmetric and 
of great help to appreciate the r\^ole of rotation.

It is easy to verify that 
the 3-metric in square brackets in Eq.~(\ref{KN}) 
corresponds to the flat Euclid space. 
One can also find  that 
$V=1+Q/(r-ia \cos\theta )$
is a {\it complex} harmonic function on $\mathbb R^3$ 
with an imaginary part being the rotation. These 
observations lead us to obtain a more general class of
solutions called an Israel-Wilson-Perj\'es (IWP) family~\cite{IWP},  
\begin{align}
\D s_4^2 = -|V|^{-2}(\D t+\omega )^2 +|V|^2 \D \vec x ^2 
\,, \qquad \vec \nabla \times \vec \omega =2{\rm Im}(\bar V\vec \nabla V)\,,
\label{IWP}
\end{align}  
where $V$ is an arbitrary complex harmonic on $\mathbb R^3$ and 
$\omega $ is obtained by quadrature.  
This enables us to generate a solution of 
the multiple Kerr-Newman objects by superposition.   
These specific supersymmetric backgrounds give us a lot of physical 
implications.  

Let us summarize here a list of 
our heuristic knowledge about supersymmetric solutions:

\underline{\it Force balance}:~
The Majumdar-Papapetrou metric~(\ref{MP}) realizes finely the force balance because
the  the Maxwell field is responsible for maintaining the black holes apart.
This is most explicitly illustrated by the complete linearity of the field equation $\vec \nabla^2 H=0$. 
The IWP solution~(\ref{IWP}) also restores the no force condition 
$\vec \nabla ^2 V=0$. Supersymmetric gravitating solutions are in mechanical
equilibrium just as those in gauge theories.

\underline{\it Zero Hawking temperature}:~
All known supersymmetric solutions describing a regular black hole  
have degenerate horizons as for the extreme Reissner-Nordstr\"om
solution. This means that the Hawking temperature~\cite{Hawking:1974sw} 
(i.e., the surface gravity of the horizon) 
vanishes. Remark that the zero temperature black holes are not always
supersymmetric, e.g.,  the extremal Kerr(-Newman) is not BPS.  
This is fairly persuasive since the inside region of the nonextremal
black holes is dynamical, which is not compatible with the mechanical
equilibrium. For instance, the Schwarzschild interior describes a
Kantowski-Sachs cosmology~\cite{Stephani:2003tm}. 
From the gravitational point of view,  degenerate horizons are 
exceptional and believed to be unrealized via the gravitational
collapse due to the cosmic censorship conjecture~\cite{Penrose:1969pc}.  
These configurations are ideal ground states of the theory. 

Since the zero-temperature black hole does not radiate 
thermal quanta, it cannot be unstable by spontaneous creation of
particles.
Moreover, the superradiant loss of charge is not permissible 
in the $N=2$ supergravity, since there are no elementary charged
particles in this theory. 
This validates the solitonic character of BPS objects.

\underline{\it No ergoregion}:~
For the Kerr-Newman metric with $M=Q$~(\ref{KN}) and the IWP solution~(\ref{IWP}), 
it is obvious that $g_{tt}<0$ is always satisfied.    
This means that the solution fails to posses an ergoregion 
even if the spacetime has nonvanishing angular momentum. It is easy to verify 
that the ergoregion does not exist in the Kerr-Newman family only when
$M=Q$.  
Thus, the asymptotically flat rotating BPS objects are free from a superradiant 
instability. That kind of a dynamical process does not occur 
in a supersymmetric background.

\underline{\it Near-horizon geometry}:~ 
In the vicinity of each point source $\vec x \sim \vec x_k$
in the Majumdar-Papapetrou solution,  
one can drop the constant term from the harmonic $H$. 
Then the Majumdar-Papapetrou metric~(\ref{MP}) is approximated by
the ${\rm AdS}_2 \times S^2$ geometry with the same curvature radii $Q_k$. 
This means that each degenerate horizon is locally isometric to the
neighbourhood of that of ${\rm AdS}_2$~\cite{Carter}.  Taking this limit, 
the novel isometry group ${\rm SO}(2,1)$ appears.  
A more relevant aspect here is that the ${\rm AdS}_2 \times S^2$ spacetime 
preserves  the maximal set of Killing spinors, i.e.,  
it is a maximally supersymmetric background to the Einstein-Maxwell
theory~\cite{KowalskiGlikman:1985im}. It follows that 
the Majumdar-Papapetrou metric connects 
two different vacua: the Minkowski spacetime at infinity and 
${\rm AdS}_2 \times S^2$ near the horizon. 
This property is also analogous to ordinary 
solitons which separate two distinct vacua.

These properties have been confirmed for 
a number of black hole systems in various (ungauged) supergravity theories. 
To reach a further conviction that these features are indeed 
universal to all supersymmetric black holes, 
a more systematic study of supersymmetric solutions 
is desirable. The above {\it ad hoc} steps or ansatz-based approaches 
are inadequate to uncover general properties of supersymmetric
solutions.

A first progression in this direction was made by Tod.~\cite{Tod} 
He assumed the existence of at least one Killing spinor $\epsilon $
satisfying Eq.~(\ref{susy}) and considered bilinears  
built out of the Killing spinor. 
These tensorial quantities fulfill many algebraic and differential constraints, which 
are sufficient to reconstruct full bosonic elements of the theory.  
The BPS solutions fall into two classes--timelike
and null family--depending on whether the Killing vector 
$V^\mu =i \bar \epsilon \gamma^\mu \epsilon $ 
constructed from the Killing spinor is timelike or null.  
The general metric of timelike class is given by the IWP family, whilst
the null family is described by the plane-fronted wave with parallel
rays~\cite{Stephani:2003tm}. 
Therefore the result in $D=4$, $N=2$ supergravity is fairly simple: the 
supersymmetric black-hole configurations are exhausted by the 
Majumdar-Papapetrou solution since the regular horizons
are destroyed by rotation.\footnote{ 
In a precise mathematical sense we need an additional technical
assumption that the Killing field that is null on the horizon is 
everywhere non-null outside the horizon~\cite{Chrusciel:2005ve}.
}
This is attributed to that the condition for the degenerate horizon  
reads $M=M(Q, J)$, whereas the Bogomol'nyi bound~(\ref{susy}) does not involve $J$. 
This property appears common to every (ungauged) supergravity theory in four
dimensions~\cite{Ortin:2008wj}.  
Additional matter sources such as dilaton~\cite{Tod2,nozawa} and 
axion ($N=4$ supergravity)~\cite{Tod2,Sen} do not produce a regular
rotating black hole in an asymptotically flat spacetime.

\section{Classification of supersymmetric solutions} 

\label{sec:classifications}

Although 
Tod's work has thrown new light on the classification 
programme of supersymmetric solutions in supergravity, 
a new technique is required in higher dimensions
since the Newman-Penrose formalism was fully used therein.
Recently, Gauntlett {\it et al.}~\cite{GGHPR} have successfully 
classified all the supersymmetric solutions of five-dimensional minimal
supergravity by making use of Killing spinor bilinears 
(see Ref.~\cite{Gst} for an early study along this line). 
This seminal work  has provided 
us with the wide range of applications.
Thereafter the classification program has achieved a remarkable development 
in diverse supergravities in various
dimensions~\cite{Gauntlett:2003fk,Gutowski:2003rg,GS5D,Caldarelli:2003pb,
GR,GP,Bellorin:2006yr,Ortin:2008wj,FigueroaO'Farrill:2002ft}.
It has been revealed that higher-dimensional supersymmetric solutions 
exhibit a considerably rich spectrum and allow 
physically interesting nontrivial black objects with rotations.

\subsection{Five dimensional minimal supergravity}

The simplest model is the minimal supergravity in five dimensions~\cite{GGHPR},
which admits the same number of supercharges as in $D=4$, $N=2$ supergravity. 
The bosonic sector of five-dimensional minimal supergravity 
constitutes the Einstein-Maxwell theory with the  Chern-Simons term.
The action is given by 
\begin{align}
S_5 =\frac{1}{16\pi G_5} \int \left({}^{(5)}R* 1 -2  F\wedge * F
 -\frac{8}{3\sqrt 3}A \wedge F \wedge F \right) \,,
\label{5D_action}
\end{align}
where $F=\D A $ is the ${\rm U}(1)$ gauge field strength, 
$\wedge$ is the wedge product 
and $*$ is the Hodge dual operator.
This theory arises via
a consistent truncation of toroidal compactification of eleven-dimensional
supergravity.  
The Chern-Simons term contributes when the solution has both of the
electric and  magnetic parts.

The strategy for classifications of supersymmetric solutions 
is parallel to Tod's argument: assume the existence 
of a Killing spinor, construct its bilinear quantities and 
derive algebraic and differential constraints which they obey.\footnote{
From a geometric standpoint, the differential forms define a 
preferred G-structure and differential conditions restrict its
intrinsic torsion (see Ref.~\cite{Gauntlett:2004hs} for a review). 
In the timelike case 
there is a local ${\rm SU}(2)$-structure whereas
in the null case we have a global $\mathbb R^3$-structure. 
Specifically, if the spacetime allows more than one Killing spinor, 
the spinorial geometry technique~\cite{spinorial_g} is more powerful than 
using the bilinears of each Killing spinor.  
}

From a pair of (commuting) symplectic-Majorana spinors 
$\epsilon^1$ and $\epsilon^2$, we can define a scalar $f$, a vector $V_M$
and three 2-forms $J^{(i)}_{MN}$ $(i=1,2,3)$ by
\begin{align}
&f=\bar \epsilon^1 \epsilon ^2 \,, \quad 
V_M =i\bar \epsilon^1 \gamma_M \epsilon^2 \,,
 \quad 
J^{(1)}_{MN} = \frac{1}{2} (\bar \epsilon^1 \gamma_{MN}\epsilon^1 +
\bar \epsilon^2 \gamma_{MN}\epsilon^2) \,, 
\nonumber \\
& \quad J^{(2)}_{MN} = -\frac{i}{2} (\bar \epsilon^1 \gamma_{MN}\epsilon^1 -
\bar \epsilon^2 \gamma_{MN}\epsilon^2) \,, 
\qquad 
J^{(3)}_{MN}= -i\bar \epsilon^1 \gamma_{MN}\epsilon^2 \,.
\label{FVJ}
\end{align}  
These quantities are not all independent: they are related to each other 
via a number of the Fierz identities. Of particular importance is
the following relation 
\begin{align}
V^MV_M=-f^2 \,,
\label{Vsquare}
\end{align} 
which implies that the vector field $V^M$ is everywhere causal. 
Other useful relations are 
\begin{align}
& J^{(i)}\wedge J^{(j)}=-2\delta^{ij}f* V \,, \quad i_V J^{(i)}=0
\,,\quad i_V* J^{(i)}=-fJ^{(i)}\,, \nonumber \\
&J_{MP}^{(i)}J_N^{(j)P}=\delta^{ij}\left(f^2g_{MN}+V_M V_N\right)
-\varepsilon_{ijk}fJ_{MN}^{(k)}\,,
\label{alg_rel}
\end{align}
where $i_V$ is the interior product and $\varepsilon_{ijk}$
is an alternating tensor.

Next, assume that the spinors $\epsilon^a$
satisfy the  Killing spinor equation
\begin{align}
 \left[\nabla_M +\frac{i}{4\sqrt 3}
\left(\gamma _{MNP}-4g_{M[N}\gamma_{P]}\right)F^{NP}\right] \epsilon^a =0
 \,. 
\label{5d_KS}
\end{align}
Differentiating the tensorial quantities~(\ref{FVJ}) and using 
the Killing spinor equation~(\ref{5d_KS}), 
we obtain the following differential constraints
\begin{align}
& \nabla_M f =\frac{2}{\sqrt 3}F_{MN}V^N\,, \qquad \nabla_MV_N
=\frac{2}{\sqrt 3}F_{MN}f+\frac{1}{2\sqrt 3}\epsilon_{MNPQR}F^{PQ}V^R
 \,, \label{diff_rel} \\
&
\nabla_M J^{(i)}_{NP}=\frac{1}{\sqrt 3}
\left[2{F_M }^Q (* J^{(i)})_{NPQ}-2{F_{[N }}^Q
 (* J^{(i)})_{P ]MQ}+g_{M [N }(* J^{(i)})_{P]QR
 }
F^{QR }\right]\,.
\nonumber
\end{align}
It then turns out that the vector field $V^M$
constructed from the  Killing spinor is a Killing field, i.e.,
$\nabla_{(M}V_{N)}=0$. One can also show that $\pounds _V F=0$ and 
$\pounds _V *F=0$
if the Bianchi identity and the Maxwell equation for $F$ hold: 
$V$ preserves all the bosonic constituents invariant.  
Consequently, the BPS solutions are naturally categorized into 
the timelike and null classes depending on the causal nature of the
supersymmetric Killing vector.

\subsubsection{Timelike family}

Let us first describe the timelike family to which all black objects with
compact horizons belong. We can take 
the vector field $V^M$ 
as a coordinate vector $V=\partial/\partial t$, for which the 
metric can be written locally as a $t$-independent form,  
\begin{align}
\D s_5^2 =-f^2 (\D t+\omega )^2 +f^{-1}h_{mn}\D x^m \D x^n \,,
\label{5D_timelike}
\end{align} 
where $f^{-1}h_{mn}$ is the metric orthogonal to the orbits of $V$. 
Indices $m,n , ...$ are raised and lowered by the base space metric 
$h_{mn}$ and its inverse. 
Split $\D \omega $ into 
the self-dual and anti-self-dual
2-forms with respect to the base space metric as
\begin{align}
f \D \omega =G^++G^- \,,\quad *_hG^\pm=\pm G^\pm \,, 
\label{domega_5D}
\end{align}
where $*_h$ is a Hodge dual with respect to the base space metric $h_{mn}$.
Then the first two differential relations in Eq.~(\ref{diff_rel}) can be solved for $F$, 
giving 
\begin{align}
 F=\frac{\sqrt 3}{2}\D \left[f(\D t+\omega )\right]-\frac{G^+}{\sqrt 3} \,.
\end{align}
The Bianchi identity and the Maxwell equations for $F$ yield the 
governing equations, 
\begin{align}
\D G^+=0 \,, \qquad \Delta  f^{-1} =\frac{2}{9}G^+_{mn}G^{+mn} \,, 
\label{Poisson}
\end{align}
where $\Delta $ is the Laplace-Beltrami operator for the base space. 
In the language of the base space, the algebraic
relations~(\ref{alg_rel}) and a differential relation~(\ref{diff_rel}) 
for the three 2-forms $J^{(i)}$
read 
\begin{align}
*_h J^{(i)}=-J^{(i)}\,,\qquad J^{(i)p}_mJ^{(j)n}_p =-\delta^{ij}
{\delta_m}^n+\varepsilon _{ijk}J^{(k)n}_m \,, \qquad 
D_m J_{np}^{(i)}=0 \,,
\end{align}
where $D_m$ is the the Levi-Civit\`a connection 
of the 4-metric $h_{mn}$.
Thus, the 2-forms $J^{(i)}$ satisfying imaginary
unit quaternions are anti-self dual and covariantly constant. 
It follows that the base space is a hyper-K\"ahler manifold with 
integrable K\"ahler forms $J^{(i)}$, i.e., its holonomy group is
contained in ${\rm Sp}(1)\simeq {\rm SU}(2)$~\cite{KN_book}. 
The Einstein equations are automatically ensured by the 
integrability condition for the Killing spinor equation. 
These conditions are necessary and sufficient for supersymmetry, 
since the Killing spinor equation is solved by 
$\epsilon=f^{1/2}\epsilon _0$ with $i\gamma^0\epsilon =\epsilon $,
where $\epsilon _0$ is a covariantly constant spinor with respect to the
base space $D_m\epsilon_0 =0$.  Now any hyper-K\"ahler manifolds 
with anti-self-dual complex structures admit
covariantly constant chiral spinors satisfying
$\gamma^{1234}\epsilon_0=\epsilon_0$.

The procedure for obtaining the timelike BPS solutions is 
as follows: choose a hyper-K\"ahler manifold and give 
a closed self-dual two-form $G^+$ wherein. Solve the Poisson equation 
for $f$~(\ref{Poisson}) with an appropriate boundary condition.
$\omega $ and $G^-$ can be obtained by solving the first equation
of~(\ref{domega_5D}) and its divergence. 
These solutions preserve at least half of the supersymmetries.

Unlike the $D=4$, $N=2$ supergravity, 
one has to solve the linear Poisson equation~(\ref{Poisson}) for a given
source $G^+$
once the base is specified. Nevertheless, this is an outstanding progress. 
In particular the solutions with nonzero $G^+$ have been missed hitherto 
in a usual impromptu approach.

The general prescription explained above exposes a striking difference
between the $D=4$ and $D=5$ cases. In $D=4$, the rotation $\omega $ 
is obtained in terms of the norm of the Killing vector as Eq.~(\ref{IWP}).
In $D=5$, on the other hand, the self-dual part of $\D \omega $ 
is an input, which is independent of the norm of the Killing. 
This notable difference in structure provides rotating black holes in
$D=5$ with regular horizons.

In what follows we enumerate some specific examples.  
Since solutions with $G^-=0$ are shown
to be inevitably static~\cite{GGHPR}, 
we shall henceforth concentrate on the solutions with $G^-\ne 0$.


\bigskip\noindent
\underline{1. \bigskip\noindent{\it BMPV black hole}}~:
Consider a flat base space $h_{mn}=\delta_{mn}$
and introduce the hyperspherical coordinates
($r, \vartheta, \phi_1, \phi_2 $) by
\begin{align}
x^1+i x^2=r\cos \vartheta e^{i\phi_1}\,, \qquad 
x^3+i x^4= r\sin \vartheta e^{i\phi_2} \,.
\label{hyperspherical}
\end{align}
Choosing $G^+=0$ and picking out only the monopole terms, 
$f$ and $\omega$ are given by
\begin{align}
 f^{-1}=1+\frac{\mu }{r^2 } \,, \qquad \omega=\frac{j}{r^2}
\left(\cos^2\vartheta \D \phi_1+\sin^2\vartheta \D \phi_2 \right) \,,
\label{BMPV}
\end{align}
where 
$\mu $ and $j$ are related to the ADM mass $M$ and angular momenta 
$J_1$ and $J_2$ as 
\begin{align}
M= \frac{3\pi \mu }{4G_5} \,, \qquad J_1=J_2=-\frac{\pi j}{4G_5}\,. 
\label{BMPV_ADM}
\end{align}
The electric charge $\ma Q$ is related to the ADM mass 
via the Bogomol'nyi relation~\cite{Gibbons:1994vm,GKLTT}
\begin{align}
M =\frac{\sqrt 3}{2} \ma Q \,.
\label{5D_BPS}
\end{align}
When $|j|<\mu^{3/2}$, the spacetime~(\ref{BMPV}) 
describes an asymptotically flat black hole with spherical topology
$S^3$.  This metric was first derived by Breckenridge-Myers-Peet-Vafa
(BMPV)~\cite{Breckenridge:1996is} 
towards the microscopic derivation of black hole entropy. 
This is the first discovery of an asymptotically flat black hole 
with nonvanishing angular momentum 
compatible with supersymmetry.
Such black holes have not been found in other dimensions.\footnote{
Horowitz and Sen~\cite{Horowitz:1995tm} 
have found a stationary BPS metric in $D~ (\ge 6)$ dimensions.   
However, this metric is singular, as encountered for the IWP solution.}
We may attribute this to the fact that
the gravitational attraction $-M/r^{D-3}$ and the centrifugal repulsion 
$J^2/M^2r^2$ delicately balance in $D=5$, illustrating the 
mechanical equilibrium. 
The BMPV metric has been rediscovered in a variety of different
contexts, as we shall discuss in $\S$~\ref{sec:intersecting_branes}. 

The BMPV spacetime has equal rotations in the 
$x^1$-$x^2$ and $x^3$-$x^4$ planes. 
This is easily understood if we express $\omega $
in terms of ${\rm SU}(2)$ left-invariant 1-forms $\sigma^i_{\rm R}$
as  $\omega =j \sigma^3_{\rm R}/(2r^2)$. 
This expression manifests that the BMPV metric admits isometries 
$\mathbb R \times {\rm SU}(2)_{\rm L}\times {\rm U}(1)_{\rm R}$. Accordingly 
the particle motion and the scalar field propagation are 
Liouville-integrable and separable~\cite{GibbonsBMPV}.

One can immediately find that the horizon at $r=0$ is degenerate, 
as that in the $D=4$ case. A notable feature of the 
BMPV solution is that the angular velocities of the horizon vanish~\cite{GMT},
i.e., the stationary Killing field which generates a time-translation at
infinity is normal to the horizon. This means that the ergoregion does
not exist.  In $D=4$, the outside region of a nonextremal black hole must be 
static if the horizon is non-rotating~\cite{Sudarsky:1992ty}. 
Accordingly, this ``staticity theorem'' 
cannot be generalized straightforwardly
to the extremal case and/or higher dimensional spacetimes.
The angular momentum at infinity merely squashes the horizon, rather
than rotates it. Computing the Komar angular momentum, one envisages the
situation that the negative fraction of angular momentum is restored
in the Maxwell fields inside the horizon~\cite{GMT}.

Another prominent feature of the BMPV solution is that
closed timelike curves exist inside the horizon $r^2<0$ 
(when $|j|<\mu ^{3/2}$). This is of course beyond the realm of 
supersymmetry since the supersymmetry transformations are essentially
local, whereas the closed timelike curves are a global notion.  
For the over-rotating case $|j|>\mu^{3/2}$, no geodesics can penetrate the
horizon, thence the spacetime is geodesically-complete~\cite{GibbonsBMPV}. 
This is understood from the area of the horizon
$A_{\rm H}=2 \pi^2\sqrt{\mu^3-j^2}$, which does not make sense in the 
over-rotating case and the entropy counting is not meaningful.

It is also enlightening to look at the near-horizon geometry of a BMPV
black hole.  Taking the scaling limit $t\to t /\epsilon^2 $
and $r\to \epsilon r$ with $\epsilon\to 0$, the metric reduces to
\begin{align}
\D s_{\rm NH}^2 =-\frac{r^4}{\mu^2} \left(\D t+\frac{j}{2r^2}\sigma^3_{\rm R}
\right)^2 +\frac{\mu }{r^2 } \left(\D r^2+r^2\D \Omega_3^2 \right) \,.
\label{NHBMPV}
\end{align}
When $j=0$, this metric reduces precisely to the 
${\rm AdS}_2\times S^3$ spacetime.\footnote{
It has been demonstrated that the general near-horizon geometry of an
extremal (not necessarily supersymmetric) black
hole in $D=4, 5$ admits enhanced isometries of 
${\rm SO}(2, 1)$~\cite{Kunduri:2007vf}.} 
The near-horizon
spacetime~(\ref{NHBMPV}) is homogeneous and isomorphic 
to the coset 
${\rm SO}(2,1)\times{\rm U}(2)/[{\rm U}(1)\times{\rm U}(1)]$~\cite{GMT}.
In addition to such an enhancement of spacetime isometries, the 
supersymmetry is also enlarged. The near-horizon BMPV metric 
(\ref{NHBMPV}) is maximally supersymmetric~\cite{GGHPR}, thereby the BMPV black hole
connects two different vacua of this theory.

A key assumption for successful derivation of black hole entropy is that
black hole solutions are uniquely determined by asymptotic charges. 
Reall demonstrated that the BMPV spacetime is indeed the unique solution among
the asymptotically flat supersymmetric black holes 
with the near-horizon geometry~(\ref{NHBMPV})~\cite{Reall2002}.

\bigskip\noindent
\underline{2. \bigskip\noindent{\it Black ring}}~:
A foremost achievement of the systematic construction of BPS solutions 
in five-dimensional minimal supergravity is the discovery of a supersymmetric
black ring~\cite{Elvang:2004rt}, which has a flat base space and nonvanishing $G^+$. 
Introducing  the ring-like coordinates $(x, y)$ via
\begin{align}
 r \cos \vartheta =\frac{R\sqrt{y^2-1}}{x-y}\,, \qquad 
 r \sin \vartheta =\frac{R\sqrt{1-x^2}}{x-y}\,,
\label{ring_coordinates}
\end{align} 
where $R$ is a constant with a dimension of length,
a flat space $\mathbb R^4$ can be written as
(with the orientation $\epsilon_{x\phi_1y\phi_2}>0$) as
\begin{align}
h_{mn}\D x^m\D x^n =\frac{R^2}{(x-y)^2} \left[
\frac{\D x^2}{1-x^2} +(1-x^2)\D \phi_1^2+
\frac{\D y^2}{y^2-1}
+(y^2-1)\D \phi_2^2 
\right]\,.
\end{align}
These coordinates cover the ranges $|x|\le 1$, 
$y\le -1$ and $\phi_{1}$, $\phi_2$ have periodicity $2\pi$.  
The overall factor $(x-y)^{-2}$ implies that $x\to y\to -1$ 
corresponds to infinity. The constant $y$ section has 
$S^1\times S^2$ orbits. 
Choosing the closed self-dual 2-form by
\begin{align}
G^+=\frac{3q}{4}\left(\D x \wedge \D \phi_1+\D y \wedge \D \phi_2\right)\,,
\end{align} 
the norm $f$ and the rotation
$\omega=\omega_{\phi_1}\D \phi_1+\omega_{\phi_2} \D \phi_2$
of the supersymmetric Killing are then given by
\begin{align}
f^{-1}& =1+\frac{Q-q^2}{2R^2}(x-y)-\frac{q^2}{4R^2}(x^2-y^2) \,,
 \nonumber \\
\omega_{\phi_1} &=-\frac{q}{8 R^2}(1-x^2)\left[3Q-q^2 (3+x+y)\right] \,,
\label{ring} \\
\omega_{\phi_2} &=\frac{3}{2}q(1+y)+\frac{q}{8 R^2}(1-y^2)\left[3Q-q^2
(3+x+y)\right] \,,\nonumber
\end{align}
where $Q$ and $q (\le Q^{1/2})$ are positive constants, corresponding
respectively to the electric charge and the local dipole charge.
The gauge potential is given by
\begin{align}
 A =\frac{\sqrt 3}{2} \left[f (\D t+\omega )-\frac{q}{2}
\biggl( (1+x)\D \phi_1+
(1+y)\D \phi_2 \biggr)\right]\,.
\label{ring_A}
\end{align}
This metric describes an asymptotically flat black ring 
$S^1\times S^2$ with a Killing horizon at 
$y\to -\infty $.  The black ring is specified by three parameters:
the ADM mass and two-independent angular momenta, 
\begin{align}
 M=\frac{3\pi Q}{4G_5}\,, \quad 
J_1=\frac{\pi q}{8G_5} \left(3Q-q^2 \right)\,,
 \quad  J_2=\frac{\pi q}{8G_5}\left(6R^2+3Q-q^2\right) \,. 
\label{ADM_ring}
\end{align}
The electric charge $\ma Q$ saturates the BPS bound~(\ref{5D_BPS}).  
While the BMPV black hole has equal angular momenta, 
the black ring never admits equal spins. Thus the asymptotic
conserved charges can distinguish these black objects with a single
event horizon,  in contrast to the vacuum
case~\cite{ER}.\footnote{
In the ${\rm U}(1)^3$-supergravity, however, the black ring has 
7 parameters and  exhibits an infinite non-uniqueness~\cite{Elvang:2004ds}. 
} 

The Bekenstein-Hawking entropy of the black ring is given by~\cite{Elvang:2004rt}
\begin{align}
S =\frac{\pi ^2 q}{4G_5} \sqrt{3 \left[(Q-q^2)^2-4q^2 R^2\right]}\,, 
\end{align}
which is real and positive if we demand
$R<(Q-q^2)/(2q)$. Imposing this condition, 
no causal pathologies occur outside the horizon. 
Similar to the BMPV spacetime,  
the ergoregion does not exist, and the Hawking 
temperature and the angular velocities of the horizon
vanish, as expected. Thus the energy extracting process 
does not occur~\cite{Nozawa:2005eu}.
Returning back to the hyperspherical coordinates~(\ref{ring_coordinates}), 
the BMPV metric is recovered by setting $R=0$ with $j=q(q^2-3Q)/4$, for which 
$q$ loses the meaning as a dipole charge. 
In the opposite infinite radius limit, 
one obtains a singly rotating black string~\cite{Bena:2004wv}.

The dipole charge $q_D$ can be defined by
\begin{align}
q_D =\frac{1}{16 \pi G_5}\int _{S^2} F =\frac{\sqrt 3 q}{16 G_5}
 \,, 
\end{align}
where $S^2$ is a surface enclosing the ring. 
One may understand this as a local distribution of 
totally zero magnetic charge. In fact it contributes to the dipole
component of asymptotic expansion of the gauge field, 
in addition to the one evoked by rotation of the ring. 
The dipole charge stems from the fact that 
the electromagnetic gauge potential is {\it not} globally well-defined
outside the horizon. Indeed one finds from Eq.~(\ref{ring_A})
that the gauge potential $A$ fails to vanish at the axis $x=1$.

Taking the near-horizon limit, one obtains the direct product of the 
${\rm AdS}_3$ with radius $q$ and the two-sphere with radius $q/2$ (see 
the original paper~\cite{Elvang:2004rt} for details). 
The ${\rm AdS}_3\times S^2$ spacetime is homogeneous and  
maximally supersymmetric~\cite{GGHPR}, so
the black ring solution also interpolates two different vacua.

It seems that the black ring does not admit an extra hidden symmetry 
other than the obvious Killing vectors $\partial_t$, $\partial_{\phi_1}$
and $\partial_{\phi_2}$. The lack of an accidental additional symmetry 
makes the quantitative study of black rings quite difficult. 
In particular the non-supersymmetric neutral black rings 
are expected to suffer from an instability 
caused by the long-wavelength gravitational perturbations. 
To the contrary, the supersymmetry should stabilize the 
BPS configuration. To see these adverse interplay is of 
primary importance, but particularly challenging.

\bigskip\noindent
\underline{3. \bigskip\noindent{\it Multiple black rings}}~:
Since the black ring solution has nonvanishing source $G^+$ 
in order to sustain the ring, it is thus obscure 
whether the multiple ring solution is constructable. 
Nevertheless, 
the superposition of solutions with nonvanishing $G^+$ 
is admissible under certain conditions~\cite{GGHPR}, as we shall discuss. 
The multiple concentric rings belong to this family~\cite{Gauntlett:2004qy}.

As a hyper-K\"ahler base space, 
we consider the Gibbons-Hawking space~\cite{Gibbons:1979zt}
\begin{align}
 \D s_{\rm GH}^2 = H^{-1}\left(\D x^5+\vec \chi\cdot \D \vec x \right)^2
 +H\D \vec x^2 \,, \qquad \vec \nabla \times \vec \chi =\vec \nabla H
 \,. \label{GH}
\end{align}
Here $\partial/\partial x^5$ is a Killing field in the four-dimensional base space
 and $H$ is harmonic on
the flat 3-space $\vec \nabla^2 H=0$. Some examples for the
Gibbons-Hawking space are the flat space ($H=1$ or $H=1/|\vec x|$), 
the Taub-NUT space ($H=1+M/|\vec x|$) and the Eguchi-Hanson space
($H=1+M/|\vec x-\vec x_0|+M/|\vec x+\vec x_0|$),
where $M$ is constant and $\vec x_0$ is a constant vector.
The Killing field $\partial/\partial x^5$ leaves the three 
complex structures invariant $\pounds_{\partial /\partial x^5} J^{(i)}=0$.

Assuming that $\partial/\partial x^5$ is a Killing vector for the whole 
five-dimensional metric, the most general solution is explicitly
obtained in terms of a set of harmonics on $\mathbb R^3$. 
Writing 
$\omega =\omega_5 (\D x^5 +\omega_i\D x^i)+\omega_i\D x^i$, 
we have
\begin{align}
G^\pm =-\frac 12 A_i^\pm \left(\D x^5+\chi_j\D x^j \right)\wedge \D x^i
\mp \frac 14 \epsilon_{ijk}A_k^\pm H\D x^i\wedge \D x^j\,,
\end{align}
where 
$
\vec A^\pm:=H^{-1}f[H\vec\nabla \omega_5\mp\omega_5\vec\nabla
H\mp\vec\nabla\times \vec\omega ]
$.
The condition $\D G^+=0$ gives
\begin{align}
 \vec \nabla\times \vec A ^+ =0 \,, \qquad 
\vec \nabla \cdot \left(H\vec A^++\vec\chi \times \vec A^+\right) =0 \,. 
\end{align}
The first equation implies that there exists a scalar function $A^+$
such that $\vec A^+=\vec \nabla A^+$. Plugging this into the second
equation gives $\vec \nabla^2 (HA^+)=0$, thereby there exists a 
harmonic function $H_{A^+}$ such that $A^+=3H_{A^+} H^{-1}$. 
Substituting this into the equation for $f$~(\ref{Poisson}) gives
\begin{align}
 \vec \nabla^2 f^{-1} = \vec\nabla^2 \left(H_{A^+}^2 H^{-1}\right)\,.
\end{align}
Thus $f$ is expressed in terms of another harmonic $H_f$ as 
$f^{-1}=H_{A^+}^2 H^{-1}+H_f$. From the expression of $\vec A^+$, 
one obtains a constraint equation for $\omega_5$. The integrability 
condition of this equation gives the governing equation of $\omega_5$ as
\begin{align}
 \vec \nabla^2 \omega_5 =\vec\nabla^2 \left(H^{-2}H_{A^+}^3 +\frac 32H^{-1}
H_{A^+}H_f \right)\,,
\end{align} 
yielding 
$\omega_5= H^{-2}H_{A^+}^3 +\frac 32H^{-1}H_{A^+}H_f+H_\omega$, 
where $H_\omega$ is yet another harmonic. 
It follows that the solution with a Gibbons-Hawking space
can be specified by four harmonics $(H,H_{A^+},H_f,H_\omega)$
provided that the five-dimensional spacetime is independent of $x^5$.

The single black ring solution is of this form with a flat 
Gibbons-Hawking metric $H=1/|\vec x|$. 
Letting $h=1/|\vec x-\vec x_0|$ with
$\vec x_0 =(0,0,-R^2/4)$, the set of harmonics are
found to be
\begin{align}
H_{A^+}=-\frac{q}{2}h\,, \qquad 
H_f= 1+\frac{Q-q^2}{4}h \,, \qquad
H_\omega =\frac{3q}{4}\left( 1-\frac{R^2}{4}h\right) \,. 
\end{align} 
Thus the multiple rings are given by~\cite{Gauntlett:2004qy}
\begin{align}
H_{A^+}=-\sum _i\frac{q_i}2 h_i\,, \quad 
H_f= 1+\sum_i\frac{Q_i-q^2_i}4h_i \,, \quad
H_\omega =\frac{3}{4}\sum_iq_i\left(
1-|\vec x_i|h_i \right)\,, 
\end{align} 
where $h_i=1/|\vec x-\vec x_i|$. 
To ensure $f>0$,  $Q_i \ge q_i^2 $ has to be satisfied. 
When all $\vec x_i$ are aligned on a $z$-axis, 
it has been shown that there exist solutions free from closed timelike
curves. The solution admits a spatial ${\rm U}(1)$-symmetry generated by
$\partial/\partial \psi$. Setting $h=1$ and $\vec \chi=0$ gives a
multiple configurations of the black string. 

The solutions on the Taub-NUT space are of interest from the Kaluza-Klein
black hole~(see Ref.~\cite{BR_TaubNUT} and 
Ishihara \& Tomizawa in this supplement). 
These solutions display the property that the spacetimes
appear to be  five dimensional in the vicinity of
the horizon, whereas they appear to be four dimensional far from the
hole. On the construction of these configurations the prescription
presented here is extremely powerful.


\bigskip
\underline{4. \bigskip\noindent{\it Multi-aligned BMPV black holes}}~:
As far as a BMPV black hole is concerned, it is fairly straightforward to 
construct multiple configuration since each black hole obeys the Laplace
equation. This kind of configuration is of course intrinsic to higher dimensions.
In four dimensions the spin-spin interaction is not enough to 
hold the Kerr black holes apart, and the BPS Kerr-Newman solution 
has no horizons.

A more interesting application is the construction of a black hole
in a {\it compactified spacetime}. 
Let us consider a flat base space $(x,y,z,w)$ and put an infinite number of BMPV
black holes along the $w$-axis with the same separation $2\pi R_5$.  
Here $R_5$ denotes the compactification radius at infinity. 
These infinite array of black holes  are identified as a black hole
living on a toroidally
compactified space ($-\pi R_5\le w\le \pi R_5$)~\cite{Myers:1986rx}. 
The desired solution is described by~\cite{Maeda_Ohta_Tanabe}
\begin{align}
f^{-1} &= 1+\sum_{k=-\infty }^{+\infty } \frac{\mu _k}{r_k^2} \,,
 \nonumber \\
\omega &= \sum_{k=-\infty }^{+\infty } j_k
\left[ \frac{x\D y-y \D x}{r_k^2 }-\frac{z \D w-(w+2\pi R_5 k) \D
 z}
{r_k^2 }\right] \,,
\end{align}
where $r_k\equiv \sqrt{x^2+y^2+z^2+(w+2\pi R_5 k)^2}$. 
When we take the limit $r_k\to 0$, 
we recover each BMPV geometry with ($\mu_k$, $j_k$) being the mass and
angular momentum parameters.  
If we take the asymptotic limit $r_k\to \infty $, 
the spacetime looks like a (twisted bundle) product of $S^1$ and the 
four dimensional spacetime. 
It is argued that from four-dimensional observer it looks like to 
exceed the Kerr-Newman cosmic censorship bound if the size of the black
hole is small compared to the compactification scale~\cite{Maeda_Ohta_Tanabe}.

\bigskip
\underline{5. \bigskip\noindent{\it Black saturn}}~:
It is interesting to see whether the configuration of a black hole
surrounded by a black ring--a black saturn--realizes. 
Although this system does not fall into class of the Gibbons-Hawking, 
Bena and Warner were able to find the desired solution~\cite{Bena:2004de}, 
\begin{align}
f^{-1}& =
(f^{-1})^{(\rm BR)}+(f^{-1})^{(\rm BH)}-1 \,,
 \nonumber \\
\omega_{\phi_1} &=
\omega_{\phi_1}^{(\rm BR)}+\omega_{\phi_1}^{(\rm BH)}
+\frac{3\mu q (1-x^2)}{2 R^2(x+y)}\,, \label{saturn}\\
\omega_{\phi_2} &=
\omega _{\phi_2}^{(\rm ring)}+\omega_{\phi_2}^{(\rm BH)}
+ \frac{3\mu q (y^2-1)}{2 R^2(x+y)}\,,\nonumber
\end{align}
where ``BR''  and ``BH'' terms refer to
the corresponding ones for the black ring~(\ref{ring}) and the 
BMPV black hole~(\ref{BMPV}), respectively. 
Written in the ring-like coordinates~(\ref{ring_coordinates}), 
quantities for the latter are given by
\begin{align}
(f^{-1})^{(\rm BH)}= 1-\frac{\mu (x-y)}{R^2 (x+y)} \,, \quad
\omega_{\phi_1}^{(\rm BH)}=\frac{j (1-x^2)}{R^2 (x+y)^2} \,, \quad
\omega_{\phi_2}^{(\rm BH)}=\frac{j (y^2-1)}{R^2 (x+y)^2} \,. 
\end{align} 
The final terms of  $\omega_{\phi 1}$ and $\omega_{\phi_2}$
in~(\ref{saturn}) describe the interaction between 
the hole and the ring. Correspondingly, the angular momenta 
pick up extra contribution,\footnote{Note that this is not a rigorous
argument since the ADM charges are defined by surface integrals at
infinity~\cite{ADM}: they cannot divide into some individual pieces.
One might expect that the Komar charge evaluated on each horizon might
be a  useful measure, but this is not the case since the event horizon is
degenerate and the spacetime is not Ricci flat.  
}  
\begin{align}
J_1 =J_1^{\rm (BR)}+J_1^{\rm (BH)}+\frac{3 \pi \mu q}{4 G_5}\,, \quad 
J_2 =J_2^{\rm (BR)}+J_2^{\rm (BH)}+\frac{3 \pi \mu q}{4 G_5}\,. 
\end{align}
whereas the ADM mass is the sum of each 
$M=M^{\rm (BR)}+M^{\rm (BH)}$. 
In order to evade closed timelike curves, we demand
$R^2+\mu <(Q-q^2)^2/(2q)^2$. 
Unlike the vacuum saturn configuration~\cite{Elvang:2007rd},
there is no frame-dragging effect due to supersymmetry. 
It may be intriguing to investigate which configurations have maximum
entropy for fixed ADM mass.

\subsubsection{Null family}

The $D=4$ BPS solutions belonging to the null class are 
wave-like solutions~\cite{nozawa,Tod,Tod2,Caldarelli:2003pb}. 
In $D\ge 5$ dimensions, the black string 
solutions fall into the null family where 
$V^MV_M=-f^2=0$~\cite{GGHPR,Gauntlett:2003fk,Gutowski:2003rg,GS5D}.

Equation~(\ref{diff_rel}) implies that $V$ is hypersurface orthogonal 
($V\wedge \D V=0$) and tangent to the affinely-parametrized geodesics 
($V^N\nabla_NV^M=0$). This means that the general solution in this case is a 
plane-fronted wave (or Kundt metric) in which all the optical 
scalars for a null vector vanish. 
It follows that one can introduce coordinates such that 
\begin{align}
 V_*=- H^{-1}\D u \,, \qquad V=\frac{\partial }{\partial v} \,, 
\end{align}
where $V_*$ is a one-form dual to $V$ and $H$ is a function independent
of $v$. 
Since the closed 2-forms $J^{(i)}$ are orthogonal to $V$, 
there exist local scalars $x^i$ such that
$J^{(i)}=\D u \wedge \D x^i$. Using the freedom $y^m\to y^m (y'^n, u)$,
one can achieve $ \nabla_m x^i={\delta^i }_m$ as inferred from the last
equation of~(\ref{alg_rel}). With this choice made, 
one finds $\gamma_{mn}=\delta_{mn}$, thereby the wave front metric 
is planner. 
Hence Eq.~(\ref{alg_rel})
determines the local metric form as 
\begin{align}
 \D s^2=-H^{-1} \left(2 \D u \D v +\ma F\D u^2\right)
+H^2 \left(\D \vec x +\vec a \D u\right)^2 \,,
\end{align}
where $\ma F$ and $\vec a$ are independent of $v$.
The differential relations~(\ref{diff_rel}) can be solved for $F$, 
giving
\begin{align}
 F= -\frac{1}{2\sqrt 3 H^2}\varepsilon _{ijk}\partial_j
 \left(H^3a_k\right)\D u \wedge \D x^i -\frac{\sqrt
 3}{4}\varepsilon_{ijk}\partial_k H
\D x^i \wedge \D x^j \,.
\end{align}
The Bianchi identity ($\D F=0$) gives rise to governing equations, 
\begin{align}
\vec \nabla^2 H=0 \,, \qquad 
\partial_u \vec \nabla H=\frac 13 \vec \nabla \times
 \left[H^{-2}\vec\nabla \times \left(H^3 \vec a \right)\right]\,. 
\end{align}
It is worth commenting that the Maxwell equations
are automatically satisfied for the null family.
The remaining function $\ma F$ is obtained by the 
($+, +$) component of the Einstein equations
\begin{align}
 \vec \nabla^2 \ma F=2 H^2 D_u W_{ii} +2 HW_{(ij)}W_{(ij)}
+\frac 13HW_{[ij]}W_{[ij]},
\end{align}
where $D_u:=\partial_u-\vec a\cdot \vec \nabla $ and 
$W_{ij}:=D_u H\delta_{ij}-H\partial_j a_i$. 
These conditions are also shown to be sufficient for supersymmetry and
the general solution is the half BPS.

\bigskip\noindent
\underline{\it Black string}. 
Choose $\ma F=\vec a =0$ and $H=H(\vec x)$. 
Then the black string solution is obtained~\cite{GGHPR},
\begin{align}
 \D s_5^2=-2 H^{-1}\D u \D v+H^2\D \vec x^2 \,, \qquad 
F=-\frac{\sqrt 3}{4}\varepsilon _{ijk}\partial_k H\D x^i \D x^j \,,
\label{BS}
\end{align}
where $H=1+Q/r$.  The coordinate transformation 
$\sqrt 2 u=t-w$ and $\sqrt 2 v=t+w$ brings the metric into the familiar
form. This illustrates that the stationary Killing field
$\partial/\partial t$ of a supersymmetric solution
is not always constructed from a Killing spinor.  
It is noted that the black string~(\ref{BS}) is distinct from the one 
obtained by the infinite radius limit of a black ring.
Note also that 
the maximally supersymmetric ${\rm AdS}_3\times S^2$ spacetime arises as the
near-horizon geometry of the black string~(\ref{BS}). 
Hence ${\rm AdS}_3\times S^2$ belongs both to the timelike and null
families, since it admits different supersymmetric Killing vectors
depending on the choice of a Killing spinor.

\subsection{Five dimensional minimal gauged supergravity}
\label{sub:gauged}

We have discussed thus far the ``ungauged'' theories admitting 
asymptotically flat spacetimes. 
When a negative cosmological constant is introduced in the
action~(\ref{5D_action}) by ${}^{(5)}R\to {}^{(5)}R+12 \ell^{-2}$, 
we come to have a gauged theory. 
The minimal gauged supergravity arises via  
consistent truncation of $S^5$-compactification of $D=10$ IIB
supergravity~\cite{Cvetic:1999xp} and have attracted much attention in the context of 
the AdS/CFT correspondence~\cite{Maldacena:1997re}. 
A systematic classification similar to the ungauged case can be
done~\cite{Gauntlett:2003fk}, but the result is in  
stark contrast with the ungaged one.  

The supersymmetric vector $V^M =i\epsilon^1\gamma^M\epsilon ^2$ 
turns out to be a causal Killing field obeying Eq.~(\ref{Vsquare}). 
Consider the case where $V$ is timelike and introduce the coordinates
with $V=\partial/\partial t$.  
The resultant metric takes the same form as~(\ref{5D_timelike}) 
modulo the base space, which is now an integrable K\"ahler manifold 
with an anti-self-dual K\"ahler form $J^{(1)}$.  
The gauge field is given by 
\begin{align}
F= \frac{\sqrt 3}{2} \D \left[f (\D t+\omega )\right]-\frac{1}{\sqrt
 3}G^+-\frac{\sqrt 3}{\ell f}J^{(1)} \,.
\label{F_AdS}
\end{align}
The notations are the same as in the previous subsections. 
Once the base is specified, 
$f$ and $G^+$ are {\it algebraically} obtained via
\begin{align}
f^{-1}=-\frac{ {}^{(h)}R \ell^2}{24} \,, \qquad 
G^+_{mn}=-\frac{\ell}{2} \left(\frac{1}{2}{}^{(h)}R_{mnpq}
J^{(1)pq}-\frac{{}^{(h)}R}{4}J^{(1)}_{mn}\right) \,,
\label{f_and_Gp}
\end{align}
where ${}^{(h)}R_{mnpq}$ and ${}^{(h)}R$
are the Riemann tensor and the Ricci scalar of the base space. 
The anti-self-dual part $G^-$ can be found through the relation
\begin{align}
\Delta f^{-1}=\frac{2}{9}G^+_{mn}G^{+mn}
+\frac{1}{\ell f}G^{-mn}J^{(1)}_{mn}-{8 \over \ell^{2} f^{2}} \,.
\label{lapf_AdS}
\end{align}
The general solutions preserve at least one quarter  of 
supersymmetries.

Observing $\D G^+=3\ell^{-1} f^{-2}\D f\wedge J^{(1)}$, 
$\D f=0$ is obtained if $\D G^+ =0$. 
It follows that in the non-rotating case ($G^+=G^-=0$), 
Eq.~(\ref{lapf_AdS}) is never satisfied, i.e.,
 a supersymmetric Killing field $V$ cannot be hypersurface-orthogonal.  
It is worth noting that this does not mean that there are no 
static BPS solutions.  

To illustrate, consider the case in which the base space is an 
Einstein space. 
Then Eq.~(\ref{f_and_Gp}) implies $G^+=0$, hence 
$f={\rm constant}\equiv 1$. For the vacuum case, 
Eq.~(\ref{F_AdS}) gives that  
$\omega $ is given by $\omega =\ell^{-1} K$, where 
$K$ is the K\"ahler potential $J^{(1)}=(1/2)\D K$.  
This is consistent with~(\ref{lapf_AdS}). 
As the base space, we make an ansatz
\begin{align}
h_{mn}\D x^m \D x^n =\frac{\D r^2}{\Delta (r)} +\frac{r^2}{4}
\left[(\sigma_1^L)^2+ (\sigma_2^L)^2 +\Delta(r) (\sigma_3^L)^2 \right] \,,
\end{align}
where $\sigma^L_i$ are the right-invariant 1-forms $S^3$. 
This metric is always K\"ahler with an anti-self-dual K\"ahler form
$
J^{(1)}=\D r\we \frac r2\sigma_3^L-\frac r2\sigma_1^L\we \frac
r2\sigma_2^L=\D(\tfrac 14 r^2\sigma_3^L)
$. 
The condition for an Einstein space gives 
$\Delta (r) = 1+r^2/\ell^2 +2\mu \ell^2/r^4$, 
where $\mu $ corresponds to the mass parameter and when 
$\mu =0$  the base space reduces to the Bergmann space.  
Changing to the non-rotating frame at infinity by 
$\phi = \phi' +(2/\ell )t $, one obtains 
\begin{align}
 \D s^2 = &-\left(1+\frac{r^2 }{\ell^2 }\right)\D t^2 
+ \frac{2\mu }{r ^2 }\left(\D t +\frac \ell 2  {\sigma'_3}^L \right)^2
+\frac{\D r^2 }{\Delta (r) }
+\frac{r^2}{4} \left[(\sigma ^L_1)^2+(\sigma^L_2)^2 +({\sigma'}^L_3)^2\right] 
\,,
\label{Kerr_AdS}
\end{align}
where ${\sigma_3'}^L=\D \phi'+\cos\theta \D \psi$. This 
is the BPS limit~\cite{Cvetic:2005nc} of the Myers-Perry-AdS
metric~\cite{Hawking:1998kw}. 
The mass and the angular momenta are given by~\cite{mass_ads}
(see also Ref.~\cite{Gibbons:2004ai}
for a careful argument of conserved
quantities in asymptotically AdS spacetimes)
\begin{align}
 M= \pi \mu \,, \qquad J_1= -J_2 = \frac{\pi \mu \ell}{2}\,.
\end{align}
One immediately finds that the solution~(\ref{Kerr_AdS})
attains the Bogomol'nyi bound~\cite{Gutowski:2004ez}
\begin{align}
M- \frac{|J_1|+|J_2|}{\ell } \ge \frac{\sqrt 3}{2}|\ma Q |\,,
\label{BPS_AdS}
\end{align}
with $\ma Q=0$.  Remark that the metric~(\ref{Kerr_AdS}) 
describes a nakedly singular spacetime. Nonetheless, it is interesting that
the vacuum metric admits a nontrivial BPS solution. This is ascribed to the  
appearance of the angular momenta in the Bogomol'nyi 
inequality~(\ref{BPS_AdS}) in the gauged case~\cite{Gutowski:2004ez,Gibbons:1983aq}. 
This metric also highlights that the stationary Killing field with respect to a
static observer at infinity is not constructed from a Killing spinor. 
When the BPS condition is imposed on the nonextremal static black hole
in AdS, the solution becomes nakedly singular in arbitrary dimensions 
for a minimal theory~\cite{Coussaert:1993jp}.

The case of $\mu =0$ is the pure AdS spacetime, for which the base space 
is the Bergmann space and the supersymmetric Killing field is rotating. 
The maximally supersymmetric vacua in the gauged case are exhausted by 
AdS~\cite{Gauntlett:2003fk}.

One finds that 
the metric~(\ref{Kerr_AdS}) possesses an ergoregion for a 
Killing vector $\partial/\partial t $ which generates a usual 
time-translation at infinity, although the supersymmetric Killing field
$V=\partial/\partial t-(2/\ell )\partial/\partial \phi' $ 
is timelike everywhere (and null on the conformal boundary). 
This is a salient feature of an asymptotically
AdS spacetime, where there exist several Killing fields which are
timelike at infinity.

It is also worth commenting that the angular 
momentum is  bounded above, which 
is generic property of black holes (irrespective of supersymmetries) in arbitrary dimensional asymptotically AdS
spacetimes~\cite{Chrusciel:2006zs}. This feature is not shared by the 
asymptotically flat black holes, for example the vacuum black ring can have  
arbitrary large angular momentum for a given mass.

For a nonvanishing electromagnetic field, one can construct 
a regular black hole solution with rotations~\cite{GR,Gutowski:2004ez}.
An interesting point to note is that the two spins are not necessarily
equal in magnitude~\cite{KLR}, contrary to the asymptotically flat case.


\section{Intersecting branes}

\label{sec:intersecting_branes}

Since  string/M-theory is formulated in ten/eleven dimensions
while our world is four dimensional, we have to study 
how to obtain realistic black holes from higher dimensions.
There are two possible mechanisms to obtain lower-dimensional world   
from higher dimensions: one is the Kaluza-Klein 
compactification of extra dimensions and the other is a brane world scenario.
In the latter case, black holes are higher dimensional if their
size is small compared enough to the bulk AdS curvature.
We just observe the three dimensional cross section of such an object. 
In the context of string theory, however,  a brane world has not been well 
organized since the domain wall dynamics does not preserve
supersymmetry. Instead, the   
compactification scenario via the Ricci flat internal space 
such as a Calabi-Yau three-fold has been actively studied and
phenomenologically preferred to provide 
an $N=1$ four-dimensional world~\cite{Candelas:1985en}.
Hence we shall explore lower  (four or five) dimensional black hole 
from the viewpoint of higher (ten or eleven) dimensional
supergravity theories.

In order to construct a (supersymmetric) black hole solution, we need to 
put gravitational sources. In the case of a Majumdar-Papapetrou solution~(\ref{MP}),
each point source gives a nonzero black hole mass as well as its charge. 
In the present setting, these singular sources are naturally ascribed 
to the presence of branes. A major utility for embedding lower
dimensional black holes into the (intersecting) D-brane picture is the
successful statistical description of black hole entropy, 
which is given classically by one quarter of the horizon area. 
In appropriate units  the  brane charges take natural numbers (counting
the number of branes), so that the entropy is also discretized. 
Study of branes in supergravity has paved the way for rapid progress
in our understanding non-perturbative regime of string/M-theory.

The possible types of branes depend on supergravity theories in question. 
Unfortunately most brane solutions do not describe black holes in their
own right (the horizon candidate might be singular).
In order for a dimensionally reduced spacetime to have a regular horizon,
multiple independent charges are required, e.g.,
we need four charges for a four-dimensional black hole which is regular
on and outside the horizon. 
As a result, we have to consider intersecting branes, which are 
uniformly distributed in compactified internal space. See e.g., 
Ref.~\cite{Gauntlett:1997cv} for a nice 
review of static intersecting branes.

In this section, we summarize how to construct
rotating black holes (or compact black  objects) in four (or five)
dimensional spacetime from ten (or eleven) dimensional branes of 
supergravity theories (see Ref.~\cite{Cvetic:1996ek} for previous studies 
of rotating branes). As described in the previous section, 
the well-organized classification algorithm works in any spacetime dimensions. 
This is also the case for the  eleven-dimensional
supergravity~\cite{GP,spinorial_g}. However, it turned out that  
the presence of a single Killing spinor 
is insufficient to determine all components of the 4-form field strength.
So we shall not follow this route. 
Alternatively, we will try to solve the Einstein equations 
with an appropriate brane configuration. 
These solutions generically fail to preserve any supersymmetries.

Consider the $D$-dimensional action composed of gravity, a massless 
dilaton field $\varphi$ and an Abelian $n_A$-field strength $F_{n_A}$, 
\begin{align}
S_D=\frac{1}{16\pi G_D} \int \D ^D x 
\left[ {}^{(D)}  R-\frac{1}{2}(\nabla\varphi)^2 -\sum_A\frac{1}{2\cdot n_A!}
e^{a_A\varphi} F_{n_A}^2 \right]\,,
\end{align}
where $A$'s denote the possible types of branes in the theory.
$a_A$ controls the coupling strength between dilaton and the gauge field.
This type of action can universally describe supergravities 
in different dimensions ($D=10$ and $11$), and encompass different
sectors of bosonic contents (Neveu-Schwarz and Ramond-Ramond).

The equations of motion derived from the action are given by
\begin{eqnarray}
&&{}^{(D)} {R}_{MN}=\frac{1}{2}\partial_M\varphi\partial_N
\varphi+\sum_A{\Theta_{n_A}}_{ MN},\nonumber\\
&&\nabla^2\varphi =\sum_A\frac{a_A}{2\cdot 
n_A!}e^{a_A\varphi}F_{{n_A}}^2,\nonumber\\
&&\partial _{M _1}(\sqrt{-g} ~e^{a_A\varphi}{F_{n_A}}^{~M_1
\cdots M_{n_A}})=0
\,,
\label{eqn:eom}
\end{eqnarray}
where ${\Theta_{n_A}}_{MN}$ is the 
stress-energy tensor of the $n_A$-form, which is given by
\begin{eqnarray}
{\Theta_{n_A}}_{MN}=\frac{1}{2\cdot n_A!}
e^{a_A\varphi}\left[n_A{{F_{n_A}}_{M}}^{R \cdots
S }{F_{n_A}}_{NR\cdots S}-\frac{n_A-1}{D-2}
F_{{n_A}}^2g_{MN}\right]
\,.
\end{eqnarray}
Here  the $n_A$-form field $F_{n_A}$  is required to fulfill the 
Bianchi identity,
\begin{eqnarray}
\partial_{[M}{F_{n_A}}_{M_1\cdots M_{n_A}]}=0
\,.
\label{eqn:bianchi}
\end{eqnarray}

\subsection{Stationary intersecting branes}

Since we are interested in supersymmetric solutions and their cousins,
we assume that there exists one timelike or null Killing
vector~\cite{GP}. 
Most studies so far have restricted primarily to static
configurations admitting a hypersurface-orthogonal timelike Killing
field. In this section we shall explore the null case, where
the spacetime admits isometry of a null orbit generated by 
$\partial/\partial v$~\cite{Maeda_Tanabe}.
As for the internal space, we assume its Ricci
flatness. We suppose that the base space spanned by $\{ x^i\}$, 
which corresponds to our world space, 
is a $(d-1)$-dimensional flat Euclid 
space, which can be extended easily to any Ricci flat space
without modification.
Under these conditions, we can adopt 
the following $D$-dimensional metric form,~\cite{Maeda_Tanabe}
\begin{eqnarray}
\D s_D^2&=&2e^{2\xi} \D u 
\left(\D v+f\D u+{{\cal A}\over \sqrt{2}}\right)
+e^{2\eta}\sum _{i=1}^{d-1}(\D x^i)^2
+\sum _{\alpha=2}^{p}e^{2\zeta
_\alpha}(\D y^\alpha)^2
\,,
\end{eqnarray}
where $D=d+p$ and 
the null coordinates $u$ and $v$ are defined by
$u=-(t-y^1)/\sqrt{2}$ and $v=(t+y^1)/\sqrt{2}$. 
This metric form describes the rotation of spacetime and the traveling wave.
Assuming that the branes extend homogeneously, 
it turns out that the metric components 
$f$, ${\cal A}={\cal A}_i \D x^i$, $\xi,\eta$
and $\zeta_\alpha$ depend only on the $(d-1)$-spatial coordinates $x^i$.

Consider a  $q_A$-brane extending into the directions
$\{y^1,y^{\alpha_2},\cdots, y^{\alpha_{q_A}}\}$. 
If the $q_A$-brane is electrically charged,  
it is well known that it couples to an  $n_A=q_A+2$ form field strength. 
Suppose that $F_{n_A}$ 
is invariant under the Killing vector $\partial/\partial v$. Assuming
the Bianchi identity,   
one can express $F_{n_A}$ in terms of the electric scalar potential
$E_{A}$ and  the magnetic vector potential $B^{A}_j$ as,  
\begin{eqnarray}
F_{{n_A}}&=&\partial _jE_{A} \D x^j\wedge
 \D u\wedge \D v
\wedge  \D y^2\wedge\cdots\wedge  \D y^{q_A}
\nonumber \\
&&
+{1\over \sqrt{2}}
\partial_i B^{A}_j  \D x^i\wedge  \D x^j\wedge  \D u
\wedge  \D y^2\wedge\cdots\wedge  \D y^{q_A}
\label{e_field}
\,,
\end{eqnarray}
In the case where 
the ${}^*{q}_A$-brane generated by a ``magnetic'' charge for  
a dual ${}^*{n}_A$-field, it just amounts to performing  
a dual transformation of the $n_A$-field with a $q_A$-brane
 (${}^*{n}_A\equiv D-n_A, {}^*{q}_A\equiv {}^*{n}_A-2$). 
In this case the form field is
described by the same form as (\ref{e_field}) for the dual field 
${}^*F_{{ {n}_A}}=F_{{ {}^*{n}_A}}$.
It is treated  as another independent form field with a
different brane from $F_{{ n_A}}$.

Under these metric and gauge field ansatz, 
we proceed to solve field equations. 
Let us define
\begin{eqnarray}
H_A&=&\exp\left[-\left(
2\xi+\sum_{\alpha=\alpha_2}^{\alpha_{q_A}}
\zeta_\alpha-{1\over 2}\epsilon_A
a_A\varphi
\right)\right]\,,\\
V&=&\exp\left[
2\xi+(d-3)\eta+\sum_{\alpha=2}^{p}
\zeta_\alpha
\right]
\,,
\label{def:H_A}
\end{eqnarray}
where 
\begin{eqnarray}
\epsilon_A =\left\{\begin{array}{ll}
+1 & \mbox{$n_A$-form field ($F_{{n_A}}$)} \\
-1 & \mbox{the dual field (${}^*F_{{{n}_A}}$)}
\end{array}\right.
\,.
\end{eqnarray}
We find that 
the field equations~(\ref{eqn:eom}) for $\xi$, $\zeta_\alpha$ and
$\varphi$ are governed by elliptic-type divergence equations: 
\begin{eqnarray}
&&\partial^j \left[V\left(\partial_j \xi
-\frac{1}{2(D-2)}\sum_A(D-q_A-3)H_A^{2}
\tilde{E}_A\partial_j\tilde{E}_A \right)\right]=0
\label{el_xi},\\
&&\partial^j\left[V\left(\partial_j \zeta_\alpha
-\frac{1}{2(D-2)}\sum_A\delta_{\alpha A}H_A^{2}
\tilde{E}_A \partial_j \tilde{E}_A\right)\right]=0
\label{el_zeta},\\
&&\partial^j\left[V\left(\partial_j \varphi
+\frac{1}{2}\sum_A\epsilon_A a_A H_A^{2}
\tilde{E}_A  \partial_j\tilde{E}_A\right)\right]=0
\,.
\label{el_varphi}
\end{eqnarray}
where we have shifted the electric potentials as 
$\tilde{E}_A=E_A-E_A^{(0)}$ with  $E_A^{(0)}$ being a constant and  we
have defined
\begin{align}
\delta_{\alpha A}=\left\{
\begin{array}{ll}
D-q_A-3 & \alpha =\alpha_2 ,..., \alpha_{q_A} \\
-(q_A+1) & {\rm otherwise} 
\end{array}
\right.
\,.
\end{align}
Though the equations of motion are considerably simplified, 
it is very difficult to find general solutions to these coupled system.
Hence we impose the following special relations:
\begin{eqnarray}
&&\partial_j \xi
=\frac{1}{2(D-2)}\sum_A(D-q_A-3)H_A^{2}
\tilde{E}_A\partial_j\tilde{E}_A 
\label{def:xi}
,\\
&&\partial_j \zeta_\alpha
=\frac{1}{2(D-2)}\sum_A \delta_{\alpha A}H_A^{2}
\tilde{E}_A \partial_j \tilde{E}_A
\label{def:zeta}
,\\
&&\partial_j \varphi
=-\frac{1}{2}\sum_A\epsilon_A a_A H_A^{2}
\tilde{E}_A  \partial_j\tilde{E}_A
\label{def:varphi}
\,,
\end{eqnarray}
Eqs.~(\ref{el_xi}), (\ref{el_zeta})  and (\ref{el_varphi})
are solved trivially by these conditions, 
which relate the first-order derivatives
of variables analogous to the BPS conditions. 
Using the definition of $V$ and the above relations, 
$\partial_j \eta$ is given by
\begin{eqnarray}
\partial_j\eta
=
-\frac{1}{2(D-2)}\sum_A (q_A+1)
H_A^{2}\tilde{E}_A\partial_j\tilde{E}_A+
 \frac{1}{(d-3)}\partial_j \ln V 
\,.
\label{def:eta}
\end{eqnarray}
Furthermore the Einstein equations lead another equation for $\eta$.
From the consistency condition with Eq.~(\ref{def:eta}), 
we find the ``intersection'' rule which 
gives the crossing dimensions
$\bar{q}_{AB}$ between two $q_A$ and $q_B$ branes  as 
\begin{eqnarray}
\bar{q}_{AB}=\frac{(q_A+1)(q_B+1)}{D-2}-1
-\frac{1}{2}\epsilon_A a_A\epsilon_B a_B
\,. 
\end{eqnarray}
and the solution of ${E}_A$ as
\begin{eqnarray}
E_A=-\sqrt{\frac{2(D-2)}{\Delta_A}}\left(1-\frac{1}{H_A}\right)
\label{q-H}
\,, 
\label{eqn:potential}
\end{eqnarray}
where we have assumed that a spacetime is  asymptotically flat 
(i.e., $H_A\rightarrow 1$ as 
$r=\sqrt{\sum_i(x^i)^2}\rightarrow \infty$) in the transverse
spatial directions and the potential $E_A$
vanishes at infinity. 
Here we have chosen the gauge of $V=1$.
These intersection rules remain unaltered compared to those for
static branes~\cite{Tseytlin:1996bh}.

Inserting this into the Maxwell equations, we obtain the equation
for $H_A$ as
\begin{eqnarray}
\partial^2 H_A=0
\,.
\label{harmonic_H}
\end{eqnarray}
Therefore $H_A$ is a harmonic function on the $(d-1)$-dimensional
flat Euclid space $\mathbb{R}^{d-1}$. Thus the system is in 
mechanical equilibrium.

Equations for the Kaluza-Klein gauge field ${\cal A}_i$  
and the wave function  $f$
reduce to the following two linear differential equations:
\begin{eqnarray}
&&
\partial^j {\cal F}_{ij} =0
\,.
\label{eq:A1}
\\
&&
\partial^2 f={\beta\over 8} \prod_{A}H_{A}^{-\frac{2(D-2)}{\Delta_{A} }}
{\cal F}_{ij} ^{~2}
\,,
\label{Poisson:f}
\end{eqnarray}
where ${\cal F}_{ij}:=\partial_i {\cal A}_j-\partial_j {\cal A}_i$ 
and $\beta$ is some constant, which is fixed by the theory 
and brane configurations (see Ref.~\cite{Maeda_Tanabe}
for detail.) 
 It follows that $\ma A$
and $f$ obey a sourceless Maxwell equation and a Poisson
equation (or a Laplace equation if $\beta=0$)
 on $\mathbb R^{d-1}$, respectively.

The solution obtained in this section
is summarized as follows: 
\begin{align}
\D s_D^2 =&\prod_A H_A^{2\frac{q_A+1}{\Delta_A}}
\left[2\prod_B H_B^{-2\frac{D-2}{\Delta_B}}
\D u\left(\D  v+f \D  u+{{\cal A}\over \sqrt{2}}\right)
\right. \nonumber \\& \left. \qquad \qquad \qquad 
+\sum_{\alpha=2}^p\prod_B 
H_B^{-2\frac{\gamma_{\alpha B}}{\Delta_B}}(\D  y^\alpha)^2
+\sum_{i=1}^{d-1}(\D  x^i)^2\right],\nonumber\\
\varphi= &(D-2)\sum_A \frac{\epsilon_A a_A}{\Delta_A} \ln H_A
\,,
\label{eqn:solution}
\end{align}
where 
\begin{eqnarray}
\Delta_A&=&(q_A+1)(D-q_A-3)
+\frac{D-2}{2}a_A^2,\nonumber\\ 
\gamma_{\alpha A}&=&\delta_{\alpha A}+q_A+1~~=~~\left\{\begin{array}{ll}
D-2 & ~~~\alpha=\alpha_2,..., \alpha_{q_A} \\
0 & ~~~\mbox{otherwise}\end{array}\right. 
\,.
\end{eqnarray}
The functions $H_A$  for each
$q_A$-brane are arbitrary harmonic, while the vector potential
$B_i^A$ is chosen as
\begin{eqnarray}
B_i^A=-\tilde{E}_A {\cal A}_i
=-{\sqrt{2(D-2)\over \Delta_A}}{{\cal A}_i\over H_A}
\,,
\end{eqnarray}
which is induced by rotation from the electric part. 
Note that the one-form 
${\cal A}={\cal A}_i \D x^i$ is unique up to the 
gradient of some scalar function $\ma A \to \ma A +\D \chi$, it can be
made to vanish by absorbing into the definition of $v$ as 
$v\to v-\chi/\sqrt 2$. 
The ``wave'' function $f$  usually satisfies the Poisson equation~(\ref{Poisson:f})
with some source term originated by the ``rotation''-induced
 metric ${\cal A}_i$, 
although it can be also an arbitrary harmonic function for some specific 
configuration of branes ($\beta=0$).

\subsection{Stationary black holes from intersecting branes}

Since the critical dimension for string/M-theory
is ten/eleven,  
we need to compactify  
extra $p (=D-d)$-dimensions down to obtain  lower 
$d $-dimensional spacetime. Branes wrapped in the compact
dimensions look like pointlike objects in $d$-dimensions, 
giving rise to black hole geometries. 
Noticing 
$\sum_\alpha \delta_{\alpha A}=(d-1)(q_A+1)-2(D-2)$ 
and 
writing the $D$-dimensional metric~(\ref{eqn:solution}) as
\begin{align}
&\D s_D^2=\Omega^{-{2\over d-2}} \D \bar{s}_d^2+\Omega_1^2 
\left[\D y^1-\frac{1}{1+f}\left(f\D t
-\frac{\cal A}{2}\right)\right]^2
+\sum_{\alpha=2}^p\Omega_\alpha^2 (\D y^\alpha)^2
\,, \nonumber \\
&
\Omega = \Omega_1 \prod_{\alpha=2}^p\Omega_\alpha  \,, \quad
\Omega_1 = (1+f)^{1/2}\prod_A H_A^{\frac{q_A+3-D}{\Delta_A}} 
\,, \quad 
\Omega_\alpha = \prod_A H_A^{\frac{q_A+1-\gamma_{\alpha A}}{\Delta_A}} \,,
\end{align}
the the $d$-dimensional Einstein frame metric is given by  
\begin{eqnarray}
\D \bar{s}_d^2
&=&
-\Xi^{d-3}\left(\D t+\frac{\cal A}{2}\right)^2
+\Xi^{-1}\sum_{i=1}^{d-1}(\D x^i)^2
\label{d_metric},\nonumber\\
\Xi&\equiv &(1+f)^{-1/(d-2)}
\prod_A H_A^{-\frac{2(D-2)}{(d-2)\Delta_A}}
\,.
\label{eqn:blackholes}
\end{eqnarray}

Assuming that the $d$-dimensional spacetime is asymptotically flat,
i.e., suppose that the metric~(\ref{eqn:blackholes}) behaves asymptotically
as  
\begin{eqnarray}
H_A \rightarrow 1+{Q_H^{(A)}\over r^{d-3}}
\,,\qquad 
f \rightarrow {Q_0\over r^{d-3}}
\,,
\end{eqnarray}
we can read off the $d$-dimensional ADM mass $M$ as
\begin{eqnarray}
M ={(d-3)\pi^{d-3\over2}\over 8 G_d \Gamma\left({d-1\over2}\right)}
\left[Q_0
+\sum_A{2(D-2)\over \Delta_A}Q_H^{(A)}\right]
\,.
\label{def:ADM_mass}
\end{eqnarray}
where $\Gamma $ is a gamma function, 
$r^2=\sum_{i=1}^{d-1}(x^i)^2$,
and $G_d $ is the $d$-dimensional gravitational constant.

As an enlightening example, we present one concrete solution to the
eleven-dimensional supergravity, in which the field content is only the 
four-form ($n_A=4$), or equivalently the dual seven-form
(${}^*n_A=7$).\footnote{
The action of eleven-dimensional supergravity is modified by a Chern-Simons term 
$A_{(3)}\wedge F_{(4)}\wedge F_{(4)}$. However, it does not contribute
to the field equation in the present settings~(\ref{e_field}).} 
Thus there exist two types of branes (M2 and M5). 
Now there is no dilaton field ($a_A=0$), so that we find  
$\Delta_A=(q_A+1)(8-q_A)=18$. 
Then  the areal radius for the $d$-dimensional Einstein 
frame metric~(\ref{eqn:blackholes}) is given by
\begin{align}
 R=r\Xi ^{-1/2} =r \left[(1+f) \prod_A H_A \right]^{1/[2(d-2)]} \,.
\label{areal_R}
\end{align}
Suppose that $f$ and $H_A$ have only the 
monopole source $\propto 1/r^{d-3}$.
Since the event horizon (if any) locates at $r=0$, it turns out that
four (three) kinds of charges are required in order for the $d=4$
($d=5$) dimensional solution to have nonzero horizon area
$R|_{r=0}>0$.

Next, let us see under which conditions these four or three charge
overlapping branes are realized.  
From he crossing rule, these two branes (M2 and M5)  can  intersect if and only if 
\begin{eqnarray}
M2\cap M2\rightarrow \bar{q}_{22}=0,\quad
M2\cap M5\rightarrow \bar{q}_{25}=1,\quad
M5\cap M5\rightarrow \bar{q}_{55}=3
\,.
\end{eqnarray}
Therefore there exist   
four-dimensional  ``black'' objects with four independent branes 
or three M5 branes plus one wave, 
and  five-dimensional ``black'' objects with three independent
 M2 branes or two branes plus one wave.
The brane configuration of the last one is shown in Table \ref{intersection1}.

\begin{table}[h]
\begin{center}
\begin{tabular}{|c|c|c|c|c|c|c|c|c|c|c|}
\hline
  $t$&
$y^1$&$y^2$&$y^3$&$y^4$&$y^5$&$y^6$&$x^1$&$x^2$&$x^3$&$x^4$\\
  \hline
$\circ$ &    $\circ$ &       &    &     &      & $\circ$ &&&&\\
  $\circ$ &  $\circ$ & $\circ$ & $\circ$ & $\circ$ & $\circ$ & &&&&   \\
 $\circ$ &     $\circ$ &    &   &    &    &    &&&&         \\
\hline
\end{tabular}
 \caption{The brane configuration of
 intersecting branes for $d=5$.}
\label{intersection1}
\end{center}
\end{table}

For these brane configuration, one arrives at
$\beta=0$~\cite{Maeda_Tanabe}, thereby  
the unknown functions $H_A (A=2,5)$,
${\cal A}_i$ and $f$ all satisfy
the harmonic equations on $\mathbb R^4$.
Adopting the hyperspherical coordinates~(\ref{hyperspherical}), 
and requiring the asymptotic flatness and the
regularity at $\vartheta=0, \pi/2$, 
we obtain the general solutions as
\begin{align}
H_A & =1+\sum_{\ell=0}^\infty
\frac{h_\ell^{(A)}}{r^{2(\ell+1)}}
P_\ell(\cos 2\vartheta)
\,,~~(A=2,5)\,, \qquad
f=\sum_{\ell=0}^\infty
\frac{Q_\ell}{r^{2(\ell+1)}}
P_\ell(\cos 2\vartheta)
\label{sol:H_sphere}
\\
{\cal A}_{\phi_1}&=\sum_{m=1}^\infty {b^{(\phi_1)}_m\over r^{2m}}
F(-m,m,1,\sin^2\vartheta)
\,,\quad 
{\cal A}_{\phi_2}=\sum_{n=1}^\infty {b^{(\phi_2)}_n\over r^{2n}}
F(-n,n,1,\cos^2\vartheta)
\,,
\end{align}
where $F(\alpha,\beta,\gamma,z)$ is the Gauss's hyper geometrical function
and $h_\ell^{(A)}, b^{(\phi_1)}_m, b^{(\phi_2)}_n$ and $Q_\ell$
are arbitrary constants. 
The solution with the lowest multipole moment is given by 
\begin{eqnarray}
H_A=1+{Q_H^{(A)}\over r^2}\,, 
\quad
f={Q_0 \over r^2}\,,
\quad
{\cal A}_{\phi_1}
={J_{\phi_1}\cos^2\vartheta\over r^2}\,,
~~~~
{\cal A}_{\phi_2}
={J_{\phi_2}\sin^2\vartheta\over r^2}
\,,
\label{BMPV3}
\end{eqnarray}
which corresponds to the (three charge generalization of) 
BMPV rotating black hole
discussed in \S~\ref{sec:classifications}.
These charges $Q_H^{(2)}$, $Q_H^{(5)}$ and $Q_0$ are interchangeable 
via U-duality~\cite{Maldacena:1996ky}. Physical properties of this
solution is quite similar to that in minimal supergravity.

The ADM mass and charges attain the Bogomol'nyi bound~\cite{Gibbons:1994vm}
\begin{eqnarray}
M &=\frac{\pi}{4G_5}\left(Q_0+Q_H^{(2)}+Q_H^{(5)}\right)
\,.
\end{eqnarray}
Although two  angular momenta
$J_\phi$ and $J_\psi$ for this solution 
are independent,
supersymmetry implies $J_{\phi_1}=J_{\phi_2}:=J$ (where we have taken
the complex structures to be anti-self-dual), in which case 
the black hole entropy is given by
\begin{eqnarray}
S=
\frac{\pi^2}{2G_5} \sqrt{
Q_0Q_H^{(2)}Q_H^{(5)}-{J^2\over 4}}
\,.
\end{eqnarray}

Adopting the other coordinate systems such as the
ring-like coordinates, 
we find that the solutions with the lowest multipole moment
give rise to naked singularities. 
As we discussed in \S~\ref{sec:classifications}, we need to include
dipole charges in order to sustain the ring. However, 
even if we take into account the source term~(\ref{Poisson:f}) with
$\beta \ne 0$,   
we failed to find a supersymmetric black ring solution
in the present framework.  
This is because we left the Chern-Simons term in eleven dimensional
supergravity out of consideration. 
If it is contained, the dimensional reduction of rotating M2/M2/M2
branes
gives rise to five-dimensional minimal supergravity coupled to
${\rm U}(1)^3$-gauge fields~\cite{Elvang:2004ds}.  
The inclusion of a Chern-Simon term in the present framework 
is an obvious next step to be argued.

\section{Dynamically intersecting branes}

\label{sec:dynamical_branes}

In this section we pay attention to the intersection of branes in a dynamical
background. 
The original idea of dynamical branes was motivated by the 
suggestion that 
colliding branes might be an alternative to conventional
four-dimensional cosmological scenarios~\cite{Khoury:2001wf}. Most studies of brane collisions 
had been done within the framework of effective field theories on the brane.
The first exact treatment of 
supergravity equations of motion was performed by Gibbons {\it et
al.}~\cite{Gibbons:2005rt}, where the authors discussed the collision of
D3-branes (see also Ref.~\cite{Chen:2005jp} for the
Ho\v{r}ava-Witten domain wall). 
These studies were restricted to the single (coincident) brane, so that
solutions do not have regular horizons. 
As described in the previous section for a stationary case, 
branes need to intersect in order to make a horizon.  It is then likely that 
we may obtain nontrivial time-dependent black objects
in four (or five) dimensions if we can extend the 
construction of intersecting branes 
described in previous section to the time-dependent settings.

\subsection{Time-dependent intersecting branes}

Since the incorporation of time dependence makes the basic equations 
difficult to be solved, 
we instead assume the same form of the metric 
as that of the static solutions:~\cite{MOU}
\begin{eqnarray}
\D s^2_D&=&-\prod_A\left[H_A(t,x)\right]^{-\frac{D-q_A-3}{D-2}}
\D t^2+\prod_A\left[H_A(t,x)\right]^{\frac{q_A+1}{D-2}}
\sum_{i=1}^{d-1}(\D x^i)^2
\nonumber \\
&&+\sum_{\alpha=1}^p\prod_A
\left[H_A(t,x)\right]^{\frac{\delta^{(\alpha)}_A}{D-2}} (\D y^\alpha)^2 
 \label{gs:metric:Eq}
\end{eqnarray}
where
\begin{eqnarray}
\delta^{(\alpha)}_A=\left\{
\begin{array}{cc}
 -(D-q_A-3)&~{\rm for}~~\alpha\parallel A\\ q_A+1 &~{\rm for}~~\alpha\perp A
\end{array} \right. \,.
\end{eqnarray}
We also assume that the scalar field $\varphi$ and
the gauge field strength $F_{n_A}$ are given by
\begin{eqnarray}
e^{\varphi}=\prod_A \left[H_A\right]^{\epsilon_A a_A/ 2}
\,,~~~F_{(q_A+2)}=\D \left[H_A^{-1}\right]
\wedge \D t\wedge \D y^{q_1}\wedge \cdots \wedge
\D y^{q_A}
\,. 
\end{eqnarray}
The only difference from the stationary case is the time-dependence of
the harmonics $H_A$.
Inserting the above ansatz into the basic equations,
we find that only one brane can have a time-dependence with a 
linear function:
\begin{eqnarray}
H_T=c_T t +\bar H_T(x)\,,~~~~H_A=H_S(x) ~~{\rm for}~~A\neq T
\,,
\label{time-dependent_brane}
\end{eqnarray} 
where $c_T$ is a constant and 
$\bar H_T(x)$ and $H_S(x)$'s are harmonics on the flat
 $(d-1)$-dimensional space. 
This type of time-dependence is too simple to describe 
realistic cosmological models. Still, it gives us a rich physical
properties and causal structures, as we will discuss below. 
Some of them have been applied  to the brane collision 
by use of multiply superposed branes.

Compactifying $D$-dimensional spacetime 
(\ref{gs:metric:Eq}) into $d$-dimensional world,
we find a time-dependent black-hole candidate solution.
Here we show one example, which is 
four-dimensional time-dependent black hole obtained from M-theory.
In M-theory (or eleven-dimensional supergravity theory), there are two
types of branes, i.e., M2 and M5.
In order to find a black hole solution in four dimensions,
we need four charges in the stationary case. 
So it is natural to consider intersections of four branes in the
time-dependent case. The configuration of M2/M2/M5/M5 brane system
is shown in Table~\ref{table:M2M2M5M5}.

\begin{table}
\begin{center}
\begin{tabular}{|c|c|c|c|c|c|c|c|c|c|c|c|}
\hline
&$t$&$y^1$&$y^2$&$y^3$&$y^4$&$y^5$&$y^6$&$y^7$&$x^1$&$x^2$&$x^3$\\
\hline
M2 & $\circ$ & $\circ$ & $\circ$ &&&&&&&& \\
\hline
M2 & $\circ$ &&& $\circ$ & $\circ$ &&&&&& \\
\hline
M5 & $\circ$ & $\circ$ &  & $\circ$ && $\circ$
& $\circ$ & $\circ$ & & &\\
\hline
M5 & $\circ$ && $\circ$ && $\circ$ & $\circ$
& $\circ$ & $\circ$ & & &\\
\hline
\end{tabular}
\end{center}
\caption{The brane configuration for M2/M2'/M5/M5', giving rise to a
 four-dimensional regular black hole.}
\label{table:M2M2M5M5}
\end{table}

The solution of the time-dependent intersecting brane system
is given by 
\begin{eqnarray}
\D s^2_{11}&=& (H_{\rm M2}H_{\rm M2'})^{-2/3} (H_{\rm M5}H_{\rm M5'})^{-1/3}\left[
      -\D t^2+H_{\rm M2}H_{\rm M2'}H_{\rm M5}H_{\rm M5'} 
\sum_{i=1}^3 (\D x^i)^2\right.
\nonumber \\
     & &\left.+H_{\rm M2'}H_{\rm M5'}(\D y^1)^2
+H_{\rm M2'}H_{\rm M5} (\D y^2)^2
+H_{\rm M2}H_{\rm M5'}(\D y^3)^2
     +H_{\rm M2}H_{\rm M5}(\D y^4)^2
     \right.
\nonumber \\
     & &\left.
     +H_{\rm M2}H_{\rm M2'} \sum_{\alpha=5}^7(\D y^\alpha)^2
\right],
\label{M2M2M5M5}\\
F_{(4)}&=&\D H_{\rm M2}^{-1}\,\wedge \D t\wedge \D y^1\wedge  \D y^2
      + \D H_{\rm M2'}^{-1}\,\wedge  \D t\wedge  \D y^3\wedge  \D y^4
\nonumber \\
       && -\ast \left( \D H_{\rm M5}\right)
        \wedge\, \D y^2\wedge  \D y^4-\ast \left( \D H_{\rm M5'}\right)
        \wedge\, \D y^1\wedge  \D y^3\,.
\end{eqnarray}
Only one of four branes (M2, M2$'$, M5, or M5$'$) is time-dependent 
and the other branes are specified by 
time-independent harmonics as Eq.~(\ref{time-dependent_brane}).
Such a time-dependent brane has a wide range of potential applications.

\subsection{Black holes in an FLRW universe}

Let us see what kind of spacetime the metric~(\ref{M2M2M5M5}) describes
upon dimensional reduction. 
Assuming that the harmonics are spatially dependent only on the radial coordinate, 
the toroidal compactification of the 
solution~(\ref{M2M2M5M5}) along $y^1$-$y^7$ produces a four-dimensional spacetime which is 
spherically symmetric, time-evolving and spatially inhomogeneous.   
The four-dimensional metric in the Einstein frame reads 
\begin{eqnarray}
\D s_4^2=-\Xi \D t^2+\Xi^{-1}\left(\D r^2+r^2 \D \Omega_2^2\right)\,,
\label{4D_metric}
\end{eqnarray}
with
\begin{eqnarray}
\Xi &=
\left[H_TH_{S}H_{S}H_{S''}
\right]^{-1/2}
\label{Xi}
\,. 
\end{eqnarray}
where
\begin{eqnarray}
H_T=\frac{t}{t_0}+\frac{Q_T}{r}\,,~~
H_S=1+\frac{Q_S}{r}\,,~~
H_{S'}=1+\frac{Q_{S'}}{r}\,,~~
H_{S''}=1+\frac{Q_{S''}}{r}
\,. 
\end{eqnarray}
The constants $Q_T$ and $Q_S, Q_{S'}, Q_{S'\hspace{-.1em}'}$ are 
charges of one time-dependent brane and three static branes,
respectively. Since all branes appear on an equal footing
in Eq.~(\ref{M2M2M5M5}), we specify 
the time-dependent branes (M2, M2$'$, M5, or M5$'$) by $T$ and
other static ones by $S$, $S'$ and $S''$.
Here and hereafter, the script ``$T$'' and ``$S$'' are understood to trace
their origin to time-dependent and static branes. 
The above metric manifests that 
the conditions of stationarity and asymptotic flatness were both relaxed. 
When all harmonics are time-independent, this solution describes nothing but an
extremely charged static black hole with an event horizon at $r=0$.

Setting all charges to zero, it is easy to find that 
the eleven-dimensional metric~(\ref{M2M2M5M5}) describes a spatially
homogeneous vacuum Kasner universe (if M2 is time-dependent the universe
contracts into $y^4$-$y^6$ directions and expands into other
directions). 
Thus we are led to a picture that
the branes are intersecting in a background of the Kasner universe. 
The fact that only one of the branes is time-dependent is compatible  
with the vacuum Einstein equations subject to the brane intersection rule.

Let us look into the asymptotic structures of the solution~\cite{MN}. 
Assuming $t/t_0>0 $ and changing to the 
new time slice $\bar t$ defined by
\begin{eqnarray}
{\bar t \over \bar t_0}
=\left ({t \over t_0} \right)^{3/4} ~~~~~{\rm with}~~~
\bar t_0=\frac{4}{3}t_0
\,,
\end{eqnarray}
the solution~(\ref{4D_metric}) can be cast 
into a more suggestive form, 
\begin{eqnarray}
\D s_4^2&=&-\bar \Xi \, \D \bar t^2+{a^2 \bar \Xi^{-1}}
\left(\D r^2+r^2 \D \Omega_2^2\right)\,,
\label{4D_metric1}
\end{eqnarray}
where 
\begin{align}
\bar \Xi &=
\left[
\left(1+{Q_T\over a^4 r}\right)
\left(1+{Q_{S}\over r}\right)
\left(1+{Q_{S'}\over r}\right)
\left(1+{Q_{S'\hspace{-.1em}'}\over r}\right)\right]^{-1/2 }
\label{Xi1}
\,, \quad 
a =\left({\bar t\over \bar t_0}\right)^{1/3} \,.
\end{align}
When we take the limit of $r\rightarrow \infty$, we can find that
the metric~(\ref{4D_metric1}) asymptotically tends to a flat FLRW spacetime, 
\begin{eqnarray}
\D s^2_4 &=&-\D \bar t^2+a(\bar t)^2
\left(\D r^2+r^2 \D \Omega_2^2\right)\,.
\label{FLRW}
\end{eqnarray}
Since the scale factor expands as $a\propto \bar t^{1/3}$, 
the universe is asymptotically filled with fluid obeying 
the stiff equation of state $P=\rho$.
On the other hand, taking the limit  $r\rightarrow 0$ with $t$ being finite,
the time-dependence turns off and the metric~(\ref{4D_metric1}) reduces to 
the ${\rm AdS}_2\times S^2$ spacetime 
with a common curvature radius 
$\ell:=(Q_TQ_{S}Q_{S'}Q_{S'\hspace{-.1em}'})^{1/4}$. 
This is a typical ``near-horizon'' geometry of an extreme black hole. 
Thus, we may speculate that this is a dynamical black hole 
with a degenerate event horizon at $r=0$ immersed in an FLRW universe.  
However, this naive picture is not true. 
Since we have fixed the time coordinate $t$ as $r\to 0$, 
the metric ${\rm AdS}_2\times S^2$ only approximates the 
``throat'' geometry. 
To identify the whole portion of horizons, 
a more rigorous treatment is required.

For simplicity 
we confine ourselves to the case of equal charges 
$Q\equiv Q_T=Q_S=Q_{S'}=Q_{S''}$ (this condition can be relaxed~\cite{MN}). 
Then it is straightforward to appreciate that 
the metric~(\ref{4D_metric1}) solves the field
equations derived from the action 
\begin{align}
S_4=\frac{1}{16\pi G}\int \D^4 x \sqrt{-g}
\left[R- (\nabla\Phi )^2 
-\sum_Ae^{\lambda_A  \Phi} F^{(A)}_{\mu\nu  } 
F^{(A)\mu\nu  }\right]\,,
\label{4Daction}
\end{align}
where $\lambda _T=\sqrt 6$ and 
$\lambda_S=\lambda_{S'}=\lambda_{S''}=-\sqrt 6/3$ are 
coupling constants. The gauge fields 
$F^{(A)}=\D A^{(A)}$ and the dilaton $\Phi$ are given by
\begin{align}
\Phi=\frac{\sqrt 6}{4}\ln \left(\frac{H_T}{H_S}\right) \,, \qquad 
A^{(T)}_t=\frac{1}{2 H_T} \,, \qquad 
A^{(S)}_t=\frac{1}{2 H_S} \,, \label{dilaton_A}
\end{align}
where 
\begin{align}
H_T=\frac{t}{t_0}+\frac{Q}{r}\,, \qquad 
H_S=1+\frac{Q}{r} \,. 
\end{align}
Namely the solution~(\ref{4D_metric}) is an exact solution in 
Einstein-dilaton-${\rm U}(1)^2$ system, which obeys the 
dominant energy condition.  Since the dilaton field $\Phi$ is
massless, it is responsible for the stiff-fluid universe.
A magnetically charged one is obtainable simply via
$\Phi\to -\Phi$ and $F^{(A)}\to e^{\lambda_A \Phi}*F^{(A)}$.

Let us examine detailed features of the solution and 
demonstrate that the solution indeed describes a black hole in an
expanding universe. 

\underline{\it Singularities}.
Inspecting~(\ref{dilaton_A}) the dilaton profile diverges
at $H_T=0$ and $H_S=0$, i.e., 
\begin{align}
 t=t_s(r):= -\frac{t_0Q}{r}\,, \qquad 
 r=-Q \,. 
\end{align}
One can verify that all the curvature invariants blow up at there. 
At these spacetime points, 
the circumferential radius $R:=r\Xi ^{-1/2}$ vanishes, 
so that they are central shell-focusing singularities. 
An investigation of null geodesics around these singularities 
implies that both singularities have the timelike structure. 
Thus the singularities are locally naked. In order to conclude whether these
singularities are covered by event horizons, we need to trace causal 
geodesics all  the way out to infinity.   
It is notable that the $t=0$ surface is no longer the big bang singularity,  
which is smoothed out due to the brane charges. One can also find that
the $r=0$ surface is not singular, hence we need to consider the region
$r<0$.

\underline{\it Trapped surfaces}. 
Since the black hole event horizon is defined by a boundary of the causal 
past of future null infinity, it is imperative to know the entire future 
evolution of spacetime in order to 
identify the locus of event horizon. From the practical point of view, 
this property is fairly awkward. Instead, it is
more advantageous to focus 
on the {\it trapped region}~\cite{HE}, on which ``outgoing'' null rays have negative 
expansion due to the strong gravitational attractive force. 
As is well known, the trapped region does not arise outside the event
horizon under the null convergence condition,  
provided that the outside of a black hole is sufficiently well-behaved~\cite{HE}. 
The outermost boundary of trapped surface in the
asymptotically flat spacetime defines the apparent horizon~\cite{HE}. 
Hayward generalized the concept of apparent horizon and introduced a 
class of {\it trapping horizons}~\cite{O21_Hayward1993}.  
One strength of the use of trapping horizons is just to 
encompass various types of horizons associated not only with black holes 
but also with white holes and cosmological ones. Hence the trapping
horizons are more  suitable in the present context. The local properties of 
 spherically symmetric spacetimes can be understood 
by the trapping properties~\cite{Nozawa:2007vq}. 
We can expect rich structures for the trapping properties since we have
two competing effects due to the black hole 
and the expanding cosmology: the former tends to focus light rays back
into the hole while the latter
tends to spread it out to infinity.

Defining null vectors
$l^{(\pm )}_\mu =\sqrt{\Xi/2}(-\nabla_\mu t \pm \Xi^{-1} \nabla_\mu r)$
orthogonal to metric sphere, 
the associated expansions are given by 
$\theta_\pm =(g_{\mu \nu }+2l^{(+)}_{(\mu }l^{(-)}_{\nu )})\nabla^\mu l^{(\pm)\nu} $.
Evaluating these expansions at the asymptotic region approximated by an FLRW universe, 
$l^{(+)}$ can be taken to be an outgoing direction.  
These expansions characterize the extent to which the light rays are
diverging or converging.  
The trapping horizons are generated by surfaces with $\theta_\pm =0$, 
which occur at $t=t^{(\mp)}_{\rm TH}(r)$, where
\begin{align}
\label{TH}
 t_{\rm TH}^{(\mp)}( r):=&
\frac{r^2}{2 t_0(H_S+3)^2}
\left[H_S^5-\frac{6t_0^2Q}{r^3}(H_S+3)
\mp H_S^3  
\sqrt{H_S^4+\frac{4t_0^2Q}{r^3}(H_S+3)}\right]\,.
\end{align}
The region $t^{(-)}_{\rm TH}<t<t^{(+)}_{\rm TH}$ with $r>0$
denotes a past trapped region of $\theta _+>0$ and $\theta_->0$  
where even ingoing null rays 
have positive expansion due to the cosmic expansion. 
Numerical calculation shows that the trapping horizon $t^{(+)}_{\rm TH}$
is spacelike for $(t_0/Q)\lesssim 5.44 $, as in the background FLRW universe. 
Whereas, the trapping horizon $t^{(-)}_{\rm TH}$ is always timelike and 
encompassing the timelike singularity $t=t_s(r)$ around which
$\theta_+<0$ and $\theta_->0$. 
The $r<0$ region can be inferred analogously. 
Just inside $r=0$, the outgoing null rays have negative expansion
$\theta_+<0$. It follows that the nature of trapped regions
$\theta_+\theta_->0$ changes considerably across $r=0$, 
which may be ascribed to the presence of a black hole. 
Thus, the $r=0$ surface of a trapping horizon might be a likely candidate of
an event horizon.

\underline{\it Near-horizon geometry}.
As we have seen, the $r=0$ limit with $t$ finite corresponds to the 
throat geometry. Inspecting the case of an extremal 
Reissner-Nordst\"om black hole, this point is not an event horizon:
the future and past event horizons are located at $r=0$ with 
$t\to \pm \infty $. In the meanwhile, 
the analysis of trapping horizon implies that  the trapping property
changes across the $r=0$ surface.
Since the trapping horizons diverge as 
$t_{\rm TH}^{(\pm)}(r) \propto \pm 1/r$ as $r\to 0$, 
the trapping horizons $t^{(+)}_{\rm TH}(r)$  and $t^{(-)}_{\rm TH}(r)$
correspond to the infinite redshift and blueshift surfaces for an
asymptotic observer. Hence the surfaces $r=0$  with $tr$ finite 
appear to be a likely candidate of the event horizons since the areal
radius remains finite in this limit. One can verify that these 
surfaces are null.  
The most suitable way 
to see the structure of these null surfaces is to take the near-horizon
limit defined by
\begin{align}
t \to {t}/{\epsilon }\,, \qquad 
r\to \epsilon r \,, \qquad \epsilon\to 0\,. 
\label{scaling} 
\end{align}
After the scaling limit, we obtain the near-horizon geometry\footnote{
Note that this ``near-horizon limit'' is not the same as the limit 
for the extremal black holes. We are just zooming up the neighbourhood
of the geometry $r=0$ and $t=\pm \infty $ with $tr$ finite. 
However, 
it can be shown that the resulting near-horizon metric~(\ref{NHmetric})
solves the field equations of the same system.   
This justifies {\it a posteriori} that the scaling limit~(\ref{scaling})
is indeed well-defined.  
}
\begin{align}
\D s_{\rm NH}^2 =- f \D t +f^{-1} \left(\D r^2+r^2 \D \Omega_{2}^2
 \right)\,, \qquad 
f =\frac{r^2}{Q^2}\left(1+\frac{tr}{t_0Q}\right)^{-1/2}\,.
\label{NHmetric}
\end{align}
The dilaton and two ${\rm U}(1)$ gauge fields also have 
a well-defined limit. 
As a direct consequence of the scaling limit~(\ref{scaling}), 
$\xi :=t\partial/\partial t-r\partial/\partial r$
is the Killing vector for the metric~(\ref{NHmetric}).
Since $\xi $ is found to be hypersurface-orthogonal in the spacetime~(\ref{NHmetric}), 
the near-horizon metric~(\ref{NHmetric}) 
can be brought into a manifestly static form,  
\begin{align}
\D s_{\rm NH}^2 
=- F(R) \D T^2 +\frac{16 (R/Q)^6}{F(R)} \D R^2 +R^2 \D \Omega_2^2 \,,
 \quad 
\end{align}
where $\xi =\partial/\partial T$, $R=rf^{-1/2}$ and 
\begin{align}
F(R) := \frac{(R^4-R_+^4)(R^4-R_-^4)}{R^2 Q^6 } \,, \quad 
R_\pm =Q\left(\frac{\sqrt{1+4\tau ^2}\pm 1}{2 \tau}\right)^{1/2}\,.
\label{Rpm}
\end{align}
Here we have defined a dimensionless parameter $\tau :=t_0/Q$. 
This is the static black hole with Killing horizons~\cite{Carter} at $R=R_\pm$, where
$\xi $ becomes null. 
It follows that the null surfaces $R=R_\pm$ in the original metric 
are the {\it nonextremal} Killing horizons, contrary to our naive
estimate.  It is also worth commenting that $\xi $ is 
not the Killing field for the original spacetime away from these
null surfaces: the outside the horizon is highly dynamical.   
Since the event horizon is described by a Killing horizon, 
the black hole fails to grow and remains the same size. 
It comes out a surprise that the 
ambient matters do not fall into the hole in spite of the 
cosmic expansion. The attractive force caused by gravity and scalar field 
cancels, on the horizon, the repulsive force of the
electromagnetic fields.

Equation~(\ref{Rpm}) gives  
$Q=\sqrt{R_+R_-}$ and $\tau =R_+R_-/(R_+^2-R_-^2)$, 
thereby the charge $Q$ sets the geometrical mean of
horizon radii and  their relative ratio is encoded in the 
dimensionless parameter $\tau$. One can also find that $\tau $ denotes the 
ratio of energy densities of electromagnetic and scalar fields at the horizon.

\begin{figure}[h]
\begin{center}
\includegraphics[width=14cm]{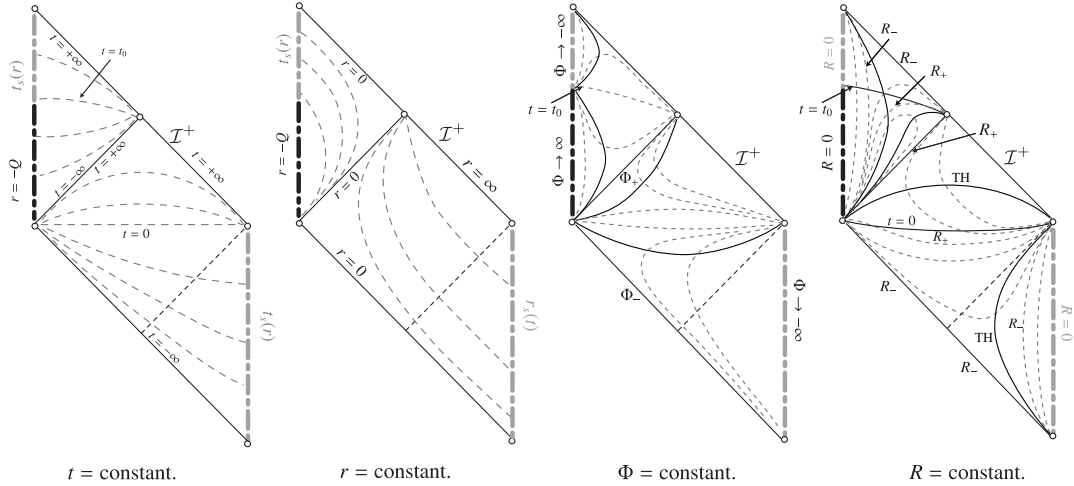}
\caption{Conformal diagram of a black hole in an expanding universe.
The central points $R=0$ describe naked singularities. 
The lines $t={\rm constant}$ and $r={\rm constant}$ are everywhere 
spacelike and timelike, respectively, while the $R={\rm constant}$ 
surfaces change signature across the trapping horizon denoted by TH. 
The $\Phi={\rm constant}$ surfaces are spacelike at infinity, according
 with  our intuition that a massless scalar field behaves like a stiff matter
 driving the cosmic expansion. The spacetime can be extended 
in a non-analytic manner across the null surfaces $R=R_-$.}
\label{PD1}
\end{center}
\end{figure}

\underline{\it Global structure.}
We are now in a position to discuss the global spacetime structure by
assembling results obtained thus far.  
Combining numerical calculations of null geodesic equations, 
we can draw the conformal diagram (Fig.~\ref{PD1}). 
Away from coordinate singularities at $r=0$ and $t=\pm \infty $, 
the $t={\rm constant}$ surface is
always spacelike, whereas the $r={\rm constant}$ surface is everywhere
timelike. 
It is shown that the solution~(\ref{4D_metric}) have a regular event horizon 
with constant radius $R_+$. 
The scalar field takes the finite values 
$\Phi_\pm=\sqrt{6}\ln (R_\pm/Q)$ at the horizons 
$R=R_\pm$.

\subsection{Generalization to arbitrary power-law expansion}

In the previous subsection, a massless scalar field drives a 
decelerating cosmic
expansion $a\propto \bar t^{1/3}$. It is then natural to ask what
happens when one includes the potential of the scalar field. 
We shall begin by the $D$-dimensional Einstein-dilaton-${\rm U}(1)^2$ system, 
in which two types of ${\rm U}(1)$-fields couple to the dilaton with different
couplings, and the dilaton has a Liouville-type exponential potential~\cite{Gibbons:2009dr,MNII}. 
To be specific, the action is described by
\begin{align}
S=& \frac{1}{16\pi G}\int \D^D x\,\sqrt{-g}\,
\left[{}^{(D)} R- \left(\nabla \Phi \right)^2
-V(\Phi)- \sum_{A=S, T}n_A 
e^{\lambda _A \Phi}
F_{\mu \nu }^{(A)} F^{(A)\mu\nu }
\right]\,,
\label{action}
\end{align}
where 
\begin{align}
V(\Phi) =V_0 \exp(-\alpha  \Phi )\,.
\label{V}
\end{align}
Here, $\alpha ~(\ge 0)$ is a dimensionless constant corresponding 
to the steepness of the potential. 
We have introduced degeneracy factors, $n_A~(\ge 0)$, 
of two ${\rm U}(1)$ fields for later convenience.
These parameters are subjected to the following relations 
\begin{align}
n_T+n_S=\frac{2(D-2)}{D-3}\,, \qquad 
\lambda_T=\alpha =-\frac{n_S}{n_T}\lambda _S= 
2\sqrt{\frac{(D-3)n_S}{(D-2)n_T}}\,.
\end{align}
The constants $n_T$ and $n_S$ may take natural number only for $D=4, 5$,
 in which
case they are related to the number of time-dependent and static branes
in eleven-dimensional supergravity~\cite{MN,MOU}.

As a natural generalization of the previous metric~(\ref{4D_metric}), 
we obtain a $D$-dimensional spatially-inhomogeneous and time-evolving metric~\cite{Gibbons:2009dr,MNII},  
\begin{align}
\D s_D^2=-\Xi^{D-3}\D t^2+\Xi ^{-1} \,
h_{mn}(x^p)\D x^m \D x^n  \,,
\label{Ddim_metric}
\end{align}
with
\begin{align}
\Xi := \left[\left(\frac{t}{t_0}+ \bar H_T \right)^{n_T}
 H_S^{n_S}\right]^{-1/(D-2)}\,,\qquad 
 t_0^2=\frac{n_T (n_T-1)}{2 V_0}\,, 
\label{Xi_p}
\end{align}
and 
\begin{align}
\Phi &=\frac{1}{2}\sqrt{\frac{(D-3)n_Tn_S}{D-2}}
\, \ln
 \left(\frac{H_T}{H_S}\right)\,,\quad 
\label{Phi}
 A^{(T)}_t=\frac{1}{2 H_T}\,,\quad 
A^{(S)}_t=\frac{1}{2 H_S}\,,
\end{align}
where  
$ \bar H_T(x)$ and $ H_S(x)$ are 
harmonics of $(D-1)$-dimensional Ricci-flat base space
$h_{mn}\D x^m \D x^n$,  and  $H_T=(t/t_0)+\bar H_T$.

When $n_T=0$, this is nothing but a static solution derived from BPS
intersecting branes (at least for $D=4, 5$). 
For $n_S=0$, the potential becomes a positive constant and 
the higher-dimensional Kastor-Traschen solution~\cite{KT,London}
is recovered. The $D=4$ and $n_T=1$ case reduces to the previous
solution with a massless scalar field derived from 
dynamically intersecting branes. For any positive values of $n_T$ and $n_S$, 
this system is shown to obey the weak energy condition.

Consider the $D=4$ case with $\bar H_T=Q/r$ 
and $H_S=1+Q/r$ for simplicity. Then the solution is specified by 
three parameters: the Maxwell charge ($Q$), the relative ratio of
energy densities of scalar and ${\rm U}(1)$-fields ($\tau= t_0/Q$) and 
the steepness parameter of the potential ($n_T$).  
The asymptotic region of the spacetime is described by the
FLRW universe~(\ref{FLRW}) with   
\begin{align}
 a(\bar t) \propto \bar t^p \,, \qquad 
 p=\frac{n_T}{4-n_T}\,.
\end{align}
Hence the parameter $n_T$ controls the expansion of the 
background cosmology:
the universe decelerates for $n_T<2$ and 
accelerates for $2<n_T\le 4$. The $n_T=4$ 
case corresponds to the exponential expansion 
caused by a cosmological constant.

\begin{figure}[t]
\begin{center}
\includegraphics[width=10cm]{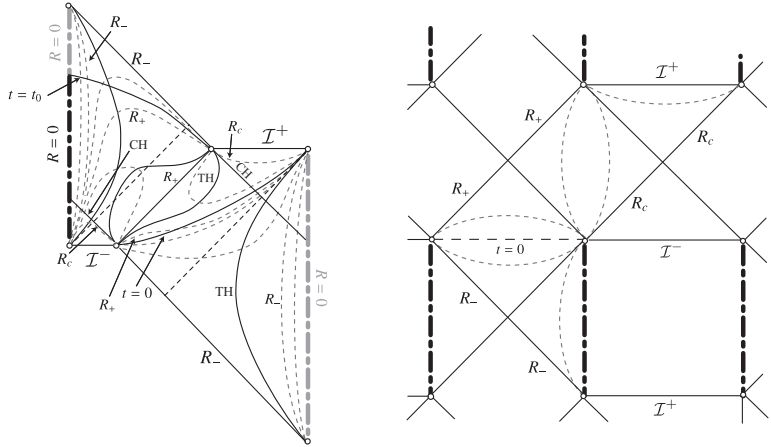}
\caption{Conformal diagrams of a black hole in an accelerating universe.
The left figure corresponds to $n_T=3$, $\tau =3$ (which is the
 representative for $2<n_T <4$ with  
$\tau \ge n_T^{n_T/2}(n_T-2)^{1-n_T/2}/2$, while the right 
is the $n_T=4$ case corresponding to the Reissner-Nordstr\"om de Sitter spacetime with $M=Q$.
In the accelerating universe, the cosmological horizon (denoted by CH)
 develops.}
\label{PD2}
\end{center}
\end{figure}

Following the argument described in the preceding 
subsection, we can obtain the conformal diagrams. 
The spacetime structures 
fall into nine types (see Table 1 in Ref.~\cite{MNII}).
Of our primary interest is the black hole in the 
accelerating universe, which arises when $n_T=4$ 
(Reissner-Nordstr\"om-de Sitter black hole with $M=Q$) or 
$2<n_T <4$ with 
$\tau \ge n_T^{n_T/2}(n_T-2)^{1-n_T/2}/2$. The conformal diagrams for
the accelerating cases are shown in Fig~\ref{PD2}.
The horizon is in general described by the nonextremal
(asymptotic) Killing horizon. A novel difference from the decelerating
universe is the existence of a cosmological horizon. 
Figure~\ref{PD1} is the representative for a black hole in the
decelerating case ($0<n_T<2$).

\subsection{Fake supergravity}

A curious property of solutions addressed in the 
preceding subsections is 
that the field equations are completely linearized.
This fact enables us to superpose the harmonics {\it ad arbitrium}
to construct multiple solutions
in spite of the time-dependence of the metric. 
This property is reminiscent of supersymmetric solution. 
But it has been well known that any dynamical phenomena are not
compatible with supersymmetry.

Nevertheless, we can in some sense give a supergravity interpretation as
follows.   
It is instrumental to consider the four-dimensional 
Einstein-Maxwell-$\Lambda$ system, where $\Lambda=-3\ell^{-2}<0$. 
This is the bosonic sector of $D=4$, $N=2$ gauged supergravity. 
The inclusion of a negative cosmological constant modifies the 
super-covariant derivative~(\ref{susy}) as 
\begin{align}
\hat \nabla_\mu \epsilon= \left(\nabla_\mu -\frac{i}{\ell}A_\mu
 +\frac{1}{2\ell}\gamma _\mu 
+\frac{i}{4}F_{\nu\rho }\gamma^{\nu\rho }\gamma_\mu \right)\epsilon 
 \,.
\label{susy_gauged} 
\end{align}
Since the (inverse of) curvature radius $\ell$ acts as a gauge coupling, 
this theory is called a gauged supergravity. Such a coupling arises 
from the R-symmetry. 
The maximally supersymmetric vacuum is only the 
anti-de Sitter space~\cite{Caldarelli:2003pb}.  
The super-covariant derivative $\hat \nabla_\mu $ is shown to be
hermite if $\ell \in \mathbb R$.

Reminding the fact that the bosonic action of $N=2$ gauged supergravity
is not charged with respect to the R-symmetry, the Wick
rotation $\ell ^{-1}\to  ih~(h\in \mathbb R)$ of a gauge coupling
amounts to changing the sign of the cosmological
constant. If there exist charged sectors, the analytic continuation
would yield unwanted ghosts~\cite{Cvetic:2004km}. 
As long as we concentrate on the truncated action composed only of
neutral fields, however,  there may appear no pathologies (at least classically).

Assuming that the Killing spinor equation $\hat \nabla_\mu \epsilon =0$
(\ref{susy_gauged}) continues to be valid after the Wick rotation
$\ell ^{-1}\to ih$, the de Sitter spacetime turns out to admit a spinor
 obeying the 1st-order
differential equation. 
Such a spinor is called a pseudo-Killing spinor and the resulting theory
is called a fake supergravity, in distinction from {\it bona fide}
supergravity.\footnote{
It has been argued in the context of fake supergravity 
that the FLRW universe is dual to supersymmetric domain walls in 
AdS~\cite{Freedman:2003ax,Skenderis:2006jq}.
The is reflected to the similarity between the 1st-order Hamilton-Jacobi equation and the 
Bogomol'nyi equation.  
}

A time-dependent pseudo-supersymmetric solution in this theory was found 
by Kastor and Traschen~\cite{KT,London}, the metric of which takes the 
exactly the same form as  Majumdar-Papapetrou solution~(\ref{MP}) up to 
the inclusion of
a linear term in time  $H\to h t +\sum_kQ_k/|\vec x-\vec x_k|$. 
This is the generalization of Majumdar-Papapetrou solution in the 
de Sitter background\footnote{Remark that the positivity proof by use
of a pseudo-Killing spinor does not
work since $\hat \nabla $ is no longer Hermite. 
In spite of the fact that the single mass Kastor-Traschen spacetime 
satisfies the analytically continued version of
the Bogomol'nyi bound $M=Q$ in AdS, 
one cannot conclude that this is the lower bound~\cite{Shiromizu:2001bg}. 
}. The Kastor-Traschen solution describes 
coalescing black holes in the contracting de Sitter universe (or
splitting white holes in the expanding de Sitter universe) and
inherits some salient characteristics from the Majumdar-Papapetrou solution. 

We can show that 
the $D=5$ solution~(\ref{Ddim_metric}) with a hyper-K\"ahler base space is in fact 
a pseudo-supersymmetric solution in minimal fake supergravity
coupled to two ${\rm U(1)}$ gauge field and a
scalar~\cite{Nozawa:2010zg}. 
It is also possible to extend to theories with arbitrary number of gauge fields
and scalars~\cite{Nozawa:2010zg,Behrndt:2003cx} and to include the
nonvanishing angular momentum.  
A notable feature of the spinning solution is that the horizon is
rotating, in contrast to the supersymmetric black holes for which the
angular velocities of the horizon vanish.   
This implies that the ergoregion exists, hence the superradiant
phenomenon occurs~\cite{Nozawa:2010zg}.
These rotating solutions generically suffer from closed  timelike curves
around the singularities. These curves may arise outside the horizon, so that
there appear naked time machines.

Recently, there has been a progress in the classifications
of pseudo-supersymmetric solutions~\cite{Grover:2008jr} and their
near-horizon geometries~\cite{Grover:2010vf}.
It is shown that the general rotating solution admits a torsion on the
hyper-K\"ahler base space. 
It is interesting to see if a black hole solution with 
nonvanishing torsion exists.

\section{Concluding remarks and outlooks}
\label{sec:conclusions}

Based on supergravity theories, which may arise from string/M-theory
in the low energy field theory limit, we have discussed supersymmetric black holes
and their relatives.
We have discussed two approaches: one is a classification method of
supersymmetric solutions in five dimensions, finding 
general black hole solutions and their relatives, 
and the other is how to construct four or five dimensional black hole solutions
by compactifying intersecting brane solutions in ten or eleven
dimensions.
These complementary strategies have unveiled very interesting solutions
describing various kinds of black objects. 
Though, the full landscape of black hole solutions, 
which is expected to be very wealthy, have been yet
uncovered. This field of study leaves much room for discussion.

For the supersymmetric black objects in supergravities, 
prime examples of unsettled open problems are the followings:

\begin{itemize}
\item It seems reasonable to anticipate that black rings exist in AdS, 
but at present no
      exact solution  is available. The most efficient way toward this is
      to impose BPS condition  as described in \S~\ref{sub:gauged}. 
It is demonstrated, however, that the black ring in minimal supergravity
      suffers from conical
singularities if it has ${\rm U}(1)^2$ spatial symmetries~\cite{Kunduri:2006uh}. 
It is likely that the black ring may be invariant under the isometry 
$V^M =i\bar \epsilon \gamma^M \epsilon $, but may not be invariant
under the usual time translation at infinity. 
Such a less symmetric black object is not excluded by the rigidity
theorem~\cite{Hollands:2006rj}, and if discovered it might trigger a rapid increase in
our knowledge about higher dimensional black holes.

\item Asymptotically flat,  
supersymmetric black holes in $D\ge 6$ have not been found yet. 
In the static case, the intersecting branes fail to 
produce supersymmetric black holes in $D\ge 6$ with nonvanishing horizon
[see Eq.~(\ref{areal_R})]. Furthermore, the solution-generating
technique is less powerful in $D\ge 6$ dimensions.  
According to the classification of Ref.~\cite{Gutowski:2003rg} for
the minimal supergravity in six dimensions, the supersymmetric Killing
field is null everywhere, which cannot be a null generator of a black hole horizon.
If we impose additional supersymmetries, the supersymmetric Killing field
may be combined to give a desired one, and things may be tractable.

\end{itemize}

In the latter case, we find the spacetime solutions from 
simple time-dependent brane solutions, which describe
black holes in the expanding universe with arbitrary power-law expansion.
Some remaining questions and future works are listed as follows:

\begin{itemize}
\item 
Some black hole solutions have a link to 
time-dependent intersecting brane systems in eleven-dimensional supergravity model.
We have not found, however,  any intersecting brane solutions in
higher dimensions, 
which reduce to the effective action involving a Liouville-type potential. 
The exponential potential is responsible for  
the expanding universe with power-law expansion.
It is interesting to find them and see whether any 
fundamental or deep reason exists. 

 \item 
Can we find more realistic black hole solutions? 
Although it was straightforward to include rotation in five
dimensions~\cite{Nozawa:2010zg}, the solution suffers from causal
       violation, hence physically unacceptable.   
A more fundamental question from the general relativistic point of view 
is whether neutral black holes exist in the expanding universe. 
When the time-dependence is incorporated for non (pseudo-)supersymmetric
solution, ambient matter will accrete into a black hole and the radius of
a black hole increases in time. These dynamical processes are desirable
       in real world,
but the construction of exact solution is a formidable task since we
       cannot resort to the 1st-order BPS equations.
The construction of these truly dynamical black holes is very hard 
under the energy condition.

\item
Can we formulate black hole thermodynamics in the time-dependent
spacetimes? In a particular case where the horizon is described by
the (asymptotic) Killing field, this is promising since the ``intensive
     parameters'' such as surface gravity, angular velocity of the
     horizon can be defined (apart from the normalization). The most
     difficult issue is how to define the mass of the hole. 
The present time-dependent solution may provide us a good tool to analyze
this question in more detail. The black hole evaporation in an expanding
universe is relevant for the observational restriction of primordial
   black holes. It is also a challenging task to account for the
 entropy from the statistical point of view. It is interesting to 
investigate if the ``attractor mechanism''~\cite{Sen:2005wa} continues to hold in a
     time-dependent case. 

\item
The black hole collision in the contracting universe may be handled 
just as the Kastor-Traschen spacetime~\cite{KT}.  
This is an outstanding  privilege since we are able to access it by
exact solution. The numerical analysis of horizon formation has
been reported in Ref.~\cite{Nakao:1994mm}.    
Similarly, it is expected that we can discuss the brane collision with
     multi time-dependent
branes, which is a generalization of the solution in Ref.~\cite{Gibbons:2005rt}.
It may be of interest to examine if the quantitative picture of coalescence changes
depending on the acceleration or the deceleration of the universe.

\item 
Although we have discussed lowest order effective theories, 
the higher curvature terms will come to be unignorable as the
      energy goes up.   It has been uncovered that 
higher curvature corrections drastically change the singularity structure
of black holes (see, e.g., papers~\cite{GB}
 and references therein).
Can we obtain exact time-dependent solutions in such theories?
A more moderate question is: how to fit the higher curvature terms  into the 
framework of (pseudo) supergravity? 
\end{itemize}

In order to promote a better understanding about higher dimensional 
(pseudo) supersymmetric black holes and their applications to real world, 
it is of crucial importance to resolve these outstanding problems.

\section*{Acknowledgements}

We would like to thank G. Gibbons, H. Maeda, U. Miyamoto, 
N. Ohta, M. Tanabe, and  K. Uzawa
for the collaboration with them, by which some results in this review
were obtained. MN is grateful to D. Klemm, R. Emparan and V. Cardoso for useful discussions
 and Waseda University for hospitality.  
This work was partially supported by the Grant-in-Aids for Scientific Research
 Fund of the JSPS (No.22540291),  
by the Waseda University Grants for Special Research Projects and 
by the MEXT Grant-in-Aid for Scientific Research on Innovative Areas No.~21111006. 

%

\end{document}